\documentclass[
reprint,
superscriptaddress,
 amsmath,amssymb,
 aps,
floatfix,
]{revtex4-2}

\usepackage{ifthen}
\usepackage{graphicx}
\usepackage{stfloats}
\usepackage{dcolumn}
\usepackage{bm}
\usepackage{hyperref}
\usepackage{float}
\usepackage{indentfirst}
\usepackage[normalem]{ulem}   
\usepackage{xcolor}

\begin{document}

\title{Electric polarization and its quantization in one-dimensional non-Hermitian chains}

\author{Jinbing Hu}
   \email{hujinbing@usst.edu.cn}
   \affiliation{College of Optical-Electrical Information and Computer Engineering, University of Shanghai for Science and Technology, Shanghai 200093, China.}
   \affiliation{Dipartimento di Fisica “Ettore Pancini”, Università degli Studi di Napoli Federico II, Complesso Universitario di Monte Sant’Angelo, Via Cintia, 80126 Napoli, Italy.}
\author{Carmine Antonio Perroni}\email{carmine.perroni@unina.it}
   \affiliation{Dipartimento di Fisica “Ettore Pancini”, Università degli Studi di Napoli Federico II, Complesso Universitario di Monte Sant’Angelo, Via Cintia, 80126 Napoli, Italy.}
\author{Giulio De Filippis}
   \affiliation{Dipartimento di Fisica “Ettore Pancini”, Università degli Studi di Napoli Federico II, Complesso Universitario di Monte Sant’Angelo, Via Cintia, 80126 Napoli, Italy.}
\author{Songlin Zhuang}
   \affiliation{College of Optical-Electrical Information and Computer Engineering, University of Shanghai for Science and Technology, Shanghai 200093, China.}
\author{Lorenzo Marrucci}
   \affiliation{Dipartimento di Fisica “Ettore Pancini”, Università degli Studi di Napoli Federico II, Complesso Universitario di Monte Sant’Angelo, Via Cintia, 80126 Napoli, Italy.}
\author{Filippo Cardano}
   \affiliation{Dipartimento di Fisica “Ettore Pancini”, Università degli Studi di Napoli Federico II, Complesso Universitario di Monte Sant’Angelo, Via Cintia, 80126 Napoli, Italy.}

\begin{abstract}
\indent We generalize the modern theory of electric polarization to the case of one-dimensional non-Hermitian systems with line-gapped spectrum. In these systems, the electronic position operator is non-Hermitian even when projected into the subspace of states below the energy gap. However, the associated Wilson-loop operator is biorthogonally unitary in the thermodynamic limit, thereby leading to real-valued electronic positions that allow for a clear definition of polarization. Non-Hermitian polarization can be quantized in the presence of certain symmetries, as for Hermitian insulators. Different from the latter case, however, in this regime polarization quantization depends also on the type of energy gap, which can be either real or imaginary, leading to a richer variety of topological phases. The most counter-intuitive example is the 1D non-Hermitian chain with time-reversal symmetry only, where non-Hermitian polarization is quantized in presence of an imaginary line-gap. We propose two specific models to provide numerical evidence supporting our findings.
\end{abstract}
\maketitle
\noindent Although the concept of electric polarization has been introduced a hundred years ago \cite{baroni1987green}, it was until the early of 1990's that the long-standing problem of crystalline polarization was solved \cite{resta1992theory,king1993theory,resta1994macroscopic,resta1998quantum,marzari2012maximally}. The main reason is that in a crystalline material the macroscopic polarization cannot be unambiguously defined as the dipole of a unit cell, since \emph{the electronic wave functions are delocalized over the whole lattice}. The first step towards a theory of polarization was made by Resta \cite{resta1992theory}, who cast the polarization difference as an integrated macroscopic current. Since then, King-Smith and Vanderbilt \cite{king1993theory} immediately built what is now known as the modern theory of polarization of crystalline insulators, which shows that the bulk polarization is strictly related to electronic geometric phases \cite{zak1989berry,berry1984quantal,xiao2010berry}. When the lattice Hamiltonian obeys to inversion and/or chiral symmetry (IS and/or CS), the polarization is quantized. Nonvanishing values of the \emph{bulk} polarization are associated with the appearance of \emph{boundary} charges \cite{benalcazar2017electric}, according to the so-called bulk-boundary correspondence (BBC) \cite{hatsugai1993chern,wang2009observation,wang2008reflection}, a hallmark of topological physics. \\
\indent When electrons in a crystalline insulator interact with the environment, their effective dynamics is described through a non-Hermitian (NH) Hamiltonian \cite{bender2002complex}. An enormous attention is currently devoted to NH physics , which is rich of unconventional phenomena like unidirectional invisibility\cite{lin2011unidirectional,regensburger2012parity,peng2014parity}, exceptional-point encirclement\cite{gao2015observation,doppler2016dynamically,yoon2018time}, enhanced sensitivity\cite{wiersig2014enhancing,liu2016metrology,lau2018fundamental}, NH skin effect \cite{yao2018edge,okuma2020topological,zhang2020correspondence,yi2020non}. These phenomena have been reproduced in a variety of artificial simulators \cite{feng2013experimental,zhen2015spawning,hodaei2017enhanced,xiao2020non}.\\
\indent Within this extremely active research field, little is known about NH electric polarization. Very recently three papers discussed NH polarization by means of many-body wave functions \cite{lee2020many}, entanglement spectrum \cite{ortega2022polarization}, and generalizing Resta's formula using biorthogonal basis\cite{Masuda2022}. Similar to the Hermitian case, these approaches considered only systems having chiral and/or inversion symmetries. However, non-Hermiticity is known to alter dramatically the definition of internal symmetries due to the distinction of complex conjugation and transposition \cite{kawabata2019symmetry,kawabata2019topological}, and to present different types of energy gap \cite{kawabata2019topological}. As such, it is crucial to understand if NH polarization is quantized in a larger variety of configurations, compared to the Hermitian case, and how these conditions are related to the type of energy gap.\\
\indent In this Letter we provide a generalization of the standard theory of electric polarization \cite{king1993theory} to line-gapped NH systems, where the bulk wave functions are extended across the entire system, like those of Hermitian crystalline materials. We follow a traditional approach, relying on the projection of the electronic position operator into the subspace of single-particle wave functions that fill the bands below the gap \cite{benalcazar2017electric}. Although the projected position operator is itself non-Hermitian, we find that it leads to a unitary Wilson-loop operator. Therefore, the Wannier centers \cite{wannier1962dynamics}, i.e. the phases of Wilson-loop eigenvalues, are purely real-valued, and so is the electric polarization, which is the summation of Wannier centers \cite{benalcazar2017electric}. Compared to other approaches \cite{lee2020many,ortega2022polarization,Masuda2022}, our derivation allows us to analyze systematically the restrictions of the basic symmetries to the Wilson-loop operator in systems with either real- or imaginary-line gaps, obtaining in turn the quantization conditions of NH polarization. With respect to Hermitian case, we find that NH systems host a larger number of topological phases, which are protected by both symmetries and the type of energy gap. The electric polarization studied here is conceptually different from the biorthogonal polarization introduced in Ref.\  \cite{kunst2018biorthogonal}, which represents a topological invariant defined in terms of zero-energy modes under open boundary conditions.\\
\indent Let us start by considering a one-dimensional (1D) NH crystalline chain composed of $N$ unit cells, each made of $N_{orb}$ lattice sites or orbitals. In this system,  the electric polarization can be computed in a single unit-cell as the dipole moment density, that is, $p=-\frac{1}{a}\sum_{\alpha=1}^{N_\text{elec}}e\, r_\alpha $, where $N_\text{elec}<N_{orb}$ is the number of electrons in each unit cell, $a$ is the lattice spacing, $e$ is the absolute value of the electron charge and $r_\alpha$'s are the electron positions with respect to the center of positive charges in the cell (see Supplementary Sec. I). For simplicity, in the remaining part of the paper we set $a=e=1$. The modern theory of electric polarization provides an elegant method to determine positions $r_\alpha$ in quantum systems where bulk electrons are delocalized, starting from the position operator projected into the subspace of occupied bands \cite{benalcazar2017electric}. A straightforward definition of the position operator is $\hat{x}=\sum_{j=1}^{N}\sum_{\alpha=1}^{N_{orb}}\hat x_{j,\alpha}$, where $\hat x_{j,\alpha}=(j+x_{\alpha}) \hat c_{j,\alpha}^{\dagger}|0\rangle\langle0|\hat c_{j,\alpha}$, with $\hat c^\dag_{j,\alpha}$ ($\hat c_{j,\alpha}$) the creation (annhilation) operators for electrons within cell $j$ and orbital $\alpha$, $x_\alpha$ the position of the orbital $\alpha$ within the unit cell, and $|0\rangle$ the vacuum state for the electrons. Unfortunately, $\hat x$ is not a legitimate operator for finite values of $N$ when periodic boundary conditions (PBC) are used. To overcome this, a unitary position operator was proposed \cite{resta1998quantum,asboth2016short}, i.e., $\hat{x}_e=\text{exp}(i2\pi\hat{x}/N)$. This operator is defined modulo $N$, thus obeying PBC. Let us note that such operator was originally discussed when considering its expectation value \cite{resta1998quantum}. However, in Ref.~\cite{de2019simple} the authors showed that this definition has to be corrected, in case one is willing to compute the distance of two coordinates, not its average value. Importantly, in the case of independent and non-interacting electrons, like our situation, both approaches give rise to the same outcome, hence we can safely use the unitary operator $\hat x_e$ defined above.\\
By discrete Fourier transformation, $\hat{x}_e$ can be alternatively written in momentum space as \cite{benalcazar2017electric}(see Sec.\ I of Supplementary Material for more detail)
\begin{equation}\label{Eq1}
    \setlength{\abovedisplayskip}{3pt}
    \setlength{\belowdisplayskip}{3pt}
    \hat{x}_e=\sum_{k,\alpha}\hat c_{k+\Delta{k},\alpha}^{\dagger}|0\rangle\langle0|\hat c_{k,\alpha},
\end{equation}
where $k\in\Delta{k}\cdot (0,1,\cdots,N-1)$, $\Delta{k}=2\pi/N$, $\hat c^\dag_{k,\alpha}$ ($\hat c_{k,\alpha}$) are creation (annhilation) operators for electrons with momentum $k$ in the orbital $\alpha$. \\
\begin{figure}[h]
    \centering
    \includegraphics[width=0.45\textwidth]{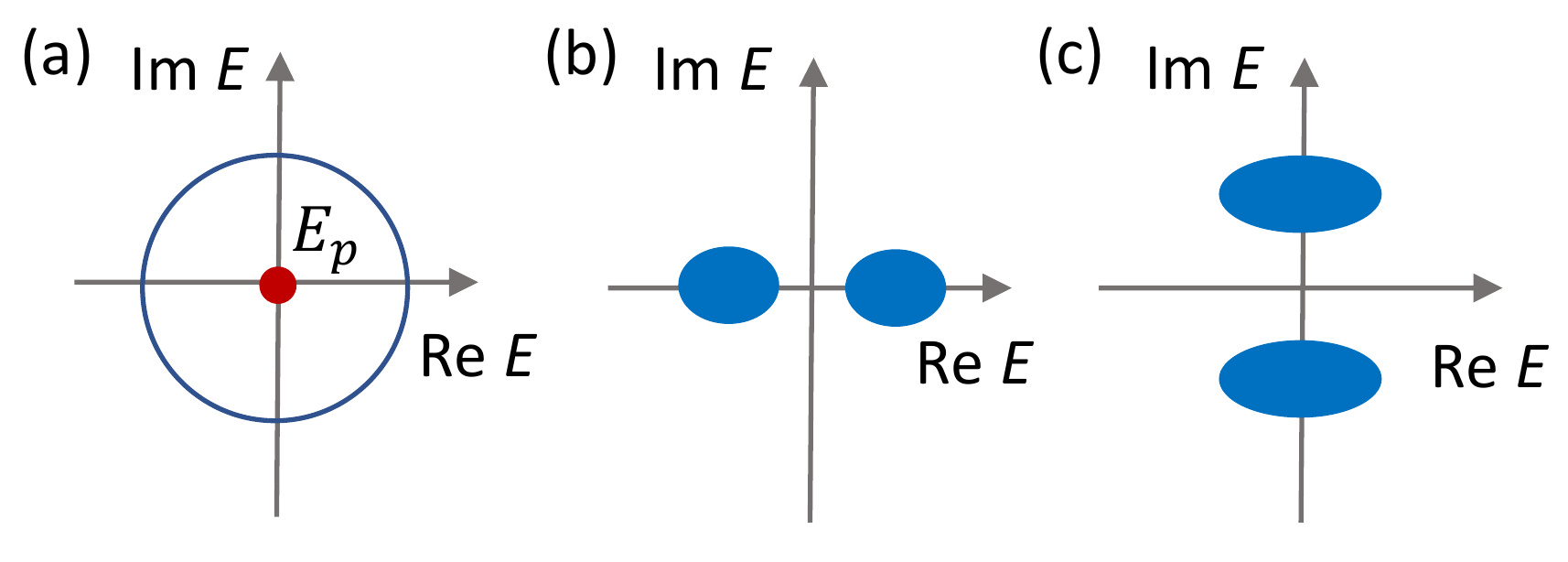}
    \caption{Possible energy gaps for NH Hamiltonians. (a) In a point-gap, there is a reference value $E_p$ that is not touched by the energy bands. (b,c) In a real/ imaginary line-gap, energies are grouped in at least two bands, whose real/imaginary part is separated by a finite gap \cite{kawabata2019symmetry}.}
    \label{Fig1}
\end{figure}
\indent Once a proper position operator has been defined, we have to consider its projection into the subspace of electronic states that are occupied  \cite{benalcazar2017electric}. Eigenstates of NH systems exhibit complex energies, hence the associated bands can exhibit different types of gaps. In particular, these can be of three kinds (see Fig.\ \ref{Fig1}), that is, a point-gap, a real line-gap and an imaginary line-gap.
The first is associated with NH systems exhibiting the NH skin effect \cite{yao2018edge,okuma2020topological,zhang2020correspondence,yi2020non}, with all eigenstates localized at the system boundaries. We here focus on line-gapped systems, whose eigenstates are delocalized across the entire chain. The definition of polarization in systems with a point-gap has been provided in Ref.\ \cite{lee2020many}, relying on a many-body formalism. In our discussion, we will consider as occupied those eigenstates whose energies, actually their real/imaginary part, are below the real/imaginary line-gap, respectively (see Fig.\ \ref{Fig1}). With this in mind, the projected position operator can be written as (see Sec. II in the Supplementary Materials for more detail):
\begin{equation}\label{Eq2}
    \setlength{\abovedisplayskip}{3pt}
    \setlength{\belowdisplayskip}{3pt}
\hat P^\text{occ}\hat{x}_e\hat P^\text{occ}=\sum_{n,m=1}^{N_\text{occ}}\sum_{k}\hat\gamma_{m,k+\Delta k}^R |0\rangle [G_k]^{mn}\langle 0|\hat\gamma_{n,k}^L,
\end{equation}
where $G_k$ is the core matrix defined in the supplementary materials, $\hat\gamma_{n,k}^j (j=R/L)$ is the right/left creation/annihilation operator of electrons with quasi-momentum $k$ in the $n$th band, and $N_\text{occ}$ is the number of bands whose energies are below the line-gap. Eq.~\ref{Eq2} is conceptually similar to Eq.~8 in Ref.\ \cite{lee2020many}, where the average of the position operator is computed for the many-body ground state. Although starting from a similar expression, here we take a different path to computing the polarization, that is, the approach harnessed in Hermitian systems \cite{benalcazar2017electric}. As we show below, this gives us the chance to obtain directly quantization conditions based on the analysis of the Wilson-loop operator.\\
\indent In the circumstance of Hermitian insulators, the projected position operator is Hermitian \cite{benalcazar2017electric}. In the NH case, however, the operator in Eq.~\ref{Eq2} is NH since it is constructed in terms of right and left eigenvectors of the NH Hamiltonian. At first sight, one would expect that this cannot generate real-valued Wannier centers. Nevertheless, by looking at the core matrix $G_k$, which is biorthogonally unitary in the thermodynamic limit, one can realize that this is not the case. By biorthogonally we mean that the core matrix is built from the biorthogonal right and left eigenvectors. By unitary, we mean that it becomes unitary in thermodynamic limit $\Delta{k}\to 0$. To see this clearly, one can perform a Taylor expansion to $[G_k]^{mn}$ and keep only terms that are linear in $\Delta{k}$ (see Sec. III in the Supplementary Materials for more details).\\
\indent On the basis of the unitary matrix $G_k$, a bi-orthogonally-unitary Wilson loop can be constructed by multiplying $G_k$ along the BZ \cite{alexandradinata2014wilson} (see Sec. IV in the Supplementary Materials for more details). As the Wilson loop operator is unitary, the phases of its eigenvalues are real numbers that still define the Wannier centers, i.e., the electronic displacement with respect to the center of positive charges. Based on this implicit relation, the NH electric polarization can be extracted as
\begin{equation}\label{Eq3}
    \setlength{\abovedisplayskip}{3pt}
    \setlength{\belowdisplayskip}{3pt}
p=\sum_{j=1}^{N_{\text{occ}}} \nu^j,
\end{equation}
where $\nu^j$ is the phase of the Wilson-loop eigenvalue, satisfying $W_{k+2\pi \gets k}|\nu_{k}^j\rangle=e^{i2\pi \nu^j}|\nu_{k}^j\rangle$, and $|\nu_k^j\rangle$ is the Wilson-loop eigenstate. In fact, the polarization defined in Eq.~\ref{Eq3} can be expressed as the integral of complex Berry connection \cite{lieu2018topological} $A_k$ over the BZ  (see Sec. V in the Supplementary Materials for derivation),
\begin{equation}\label{Eq4}
    \setlength{\abovedisplayskip}{3pt}
    \setlength{\belowdisplayskip}{3pt}
p=-\frac{1}{2\pi}\oint_{BZ}Tr[A_k]dk~\text{mod}~1,
\end{equation}
which agrees with the well-known expression of the polarization \cite{king1993theory,resta1994macroscopic,resta1998quantum}, that is, the electric polarization is a natural topological invariant. \\
\noindent\emph{Quantization conditions} $-$ As mentioned above, the unitary feature of Wilson-loop operator is crucial not only for generating real-valued electronic displacements, but also for analyzing the quantization conditions of NH polarization. In other words, Eq.~\ref{Eq3} provides an alternative way to determine the quantization conditions of NH polarization by analyzing the restriction of symmetries to the unitary Wilson-loop operator.
Symmetries play a pivotal role in quantizing electric polarization, IS for instance forces the polarization to be either 0 or 1/2 in Hermitian crystalline insulators \cite{benalcazar2017electric}. In NH physics, symmetries are dramatically altered due to the distinction of complex conjugation and transposition \cite{kawabata2019symmetry,yi2020non}. Generally, there are seven basic symmetries, i.e., IS, CS, anomalous CS (CS$^{\dagger}$), TRS, anomalous TRS (TRS$^{\dagger}$), PHS, and anomalous PHS (PHS$^{\dagger}$), with related operators $I, \Gamma, S, T, \tilde{T}, C, \tilde{C}$, respectively (see Sect. VI of Supplementary Materials for explicit definitions). Going beyond previous studies, here we investigate the constraint of these basic symmetries to NH polarization, both in real and imaginary gapped systems, which are associated with different symmetry-protected topological phases \cite{kawabata2019symmetry}.\\
\begin{table}[H]
    \caption{The quantization table of NH electric polarization for seven basic symmetries. ‘$r$’ and ‘$i$’ denote real-line and imaginary-line gap, respectively, and $\surd$ indicates the quantization of bulk polarization by the corresponding symmetry. Note, for each of the last three types of symmetries, the quantization of polarization is only applicable when the symmetry operator g satisfies $gg^*=1~(g=T, C, \tilde{C})$.}
    \begin{ruledtabular}
    \begin{tabular}{cccccccc}
    \textrm{}&\textrm{IS}&\textrm{CS}&\textrm{CS$^{\dagger}$}&\textrm{TRS$^{\dagger}$}&\textrm{TRS}&\textrm{PHS}&\textrm{PHS$^{\dagger}$}\\
    \hline
    $r$ & $\surd$ & $\surd$ & $\surd$ & &         & $\surd$ & $\surd$\\
    $i$ & $\surd$ &         & $\surd$ & & $\surd$ & $\surd$ &        \\
    \end{tabular}
    \label{Table1}
    \end{ruledtabular}
\end{table}
\indent The results we obtain are summarized in Table.~\ref{Table1} (see Secs. VI and VII of Supplementary Materials for concrete derivation), which clearly shows that, compared to the Hermitian case, NH polarization is quantized in a larger number of symmetry classes. In general, the quantization conditions can be categorized into three classes: the first includes TRS$^{\dagger}$ only, imposing no restriction to the NH polarization; the second class includes IS, CS$^{\dagger}$,and PHS, with each presenting quantized bulk-polarization for both real- and imaginary-line gaps; the last class includes CS, TRS, PHS$^{\dagger}$, which quantizes NH polarization only for real- or imaginary-line gap. Note that, for each of the last three symmetries in Table.~\ref{Table1}, i.e., TRS, PHS, PHS$^{\dagger}$, the quantization condition is only applicable when the symmetry operator $g$ satisfies $gg^*=1(g=T, C, \tilde{C})$, while the unitary sewing matrix is zero for $gg^*=-1(g=T, C, \tilde{C})$ (see subsection 4 of Sec. VI in the Supplementary Materials for more details). \\
\indent Among the cases summarized in Table.~\ref{Table1}, systems with TRS and imaginary-line gap represent a significant example, which is fully distinct with respect to the Hermitian case. In Hermitian insulators indeed, TRS imposes no restrictions to electric polarization because the energies satisfy the relation $E_{\pm}(k)=E_{\pm}(-k)$, and bands are individually subject to TRS, thus, the topological phase is absent \cite{kawabata2019topological}. In NH physics, however, under the constraint of TRS, the energy spectrum satisfies the relation $E_+(k)=E^*_-(-k)$. If the energy spectrum is characterized by an imaginary-line gap, then the positive and negative imaginary bands are always paired; considering the fact that the Hilbert space of all bands is topologically trivial \cite{benalcazar2017electric}, then topological phase featured with $\mathbb{Z}_2$ invariant appears \cite{kawabata2019symmetry}. Note that the quantization mechanism of TRS associated with imaginary-line gap is the same as that of PHS$^{\dagger}$ associated with real-line gap, due to the unification of TRS and PHS$^{\dagger}$ \cite{kawabata2019topological}.\\
\begin{figure}[t]
    \centering
    \includegraphics[width=0.45\textwidth]{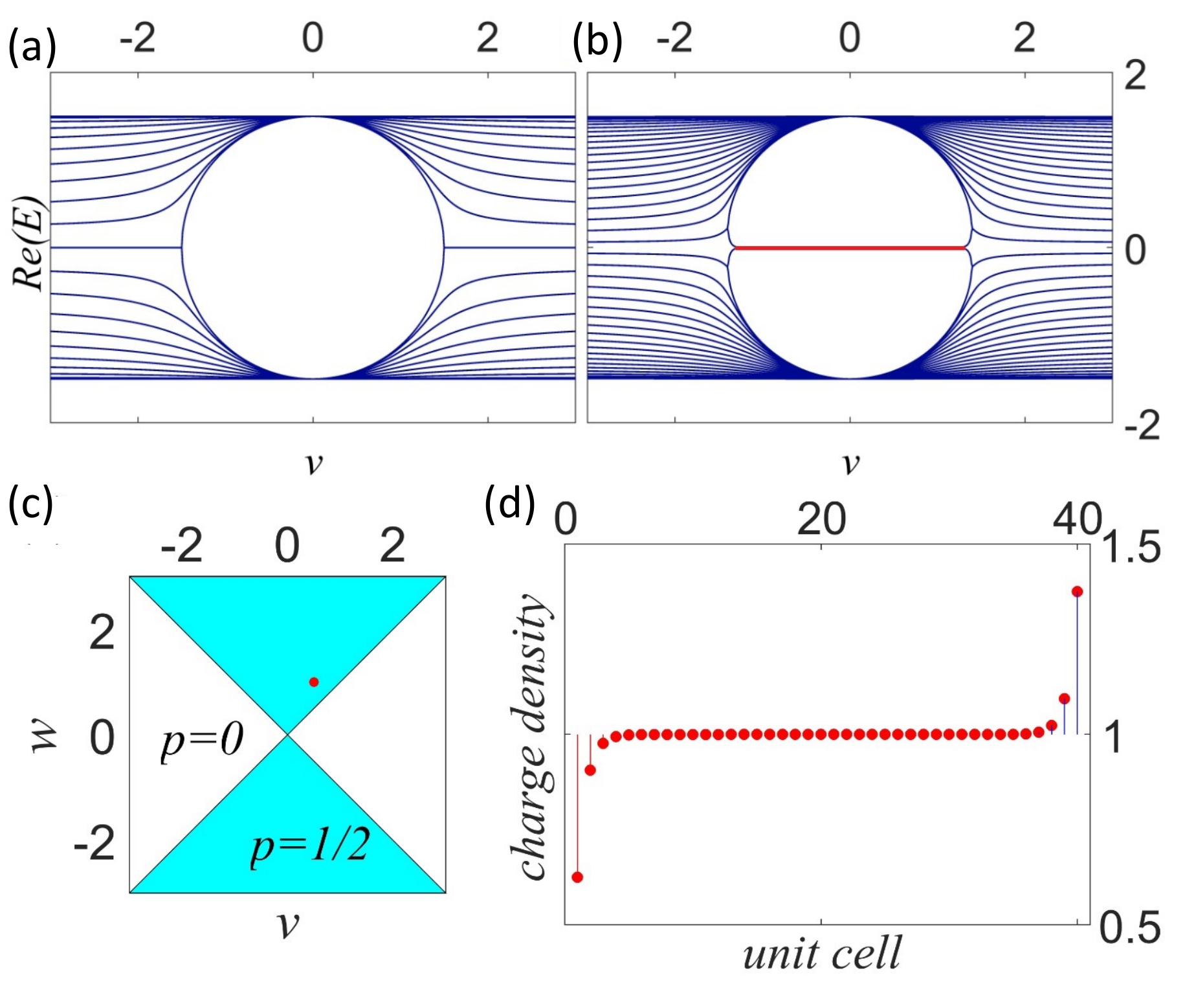}
    \caption{The real part of the eigenvalues of the NH Hamiltonian defined in Eq.~\ref{Eq5}, when $w=1.5$, $N=40$, and $v\in(-3,3)$, obtained for (a) periodic and (b) open boundary conditions, respectively. The red line in (b) marks the zero-energy edge modes. (c) The polarization phase diagram as functions of $w,v\in(-3,3)$. (d) Electron charge distribution in the topologically nontrivial polarization phase for $w = 1.0, v = 0.5$ (i.e., the red dot denoted in (c)). The total electron charge at the ends is $\pm e/2$ relative to background.}
    \label{Fig2}
\end{figure}
\noindent\emph{Specific Examples}$-$Here we present two simple models (with two sites per unit cell) to provide numerical evidence for some of the results obtained above. First, let us consider a bi-partite lattice where electrons are subject to a NH Su-Schrieffer-Heeger (SSH) Hamiltonian, that exhibits NH polarization due to the presence of CS$^{\dagger}$. The Hamiltonian reads
\begin{equation}\label{Eq5}
    \setlength{\abovedisplayskip}{3pt}
    \setlength{\belowdisplayskip}{3pt}
\hat{H}=\sum_{j}[v(\hat{a}_j^{\dagger}\hat{b}_j-\hat{b}_j^{\dagger}\hat{a}_j)+w(\hat{a}_j^{\dagger}\hat{b}_{j-1}+\hat{b}_{j-1}^{\dagger}\hat{a}_j)],
\end{equation}
where $\hat{a}_j (\hat{a}_j^{\dagger})$ and $\hat{b}_j (\hat{b}_j^{\dagger})$ are annihilation (creation) operators on the sites ‘A’ and ‘B’ of $j$-th cell, respectively, and $v,w\in\mathbb{R}$ denote hopping amplitudes. In this model, the non-Hermiticity is induced through the negative sign of the intracell coupling. The Hamiltonian in Eq.~\ref{Eq5} obeys to CS$^{\dagger}$ and TRS, with the corresponding operator $S=\oplus_{j=1}^{N}\sigma_z$ and $T=\oplus_{j=1}^N\sigma_0$, where $\sigma_0$ and $\sigma_z$ are identity matrix and Pauli matrix, respectively. These symmetries force the eigenenergies to come in $(E,-E^*)$ pairs, thus presenting a real-line gap in the energy spectrum (see Fig.~\ref{Fig2}(a,b), the complete spectrum is in Sect. VIII of the SM). By means of the NH polarization formula reported in Eq.~\ref{Eq3}, the topological phase diagram is derived as functions of $w$ and $v$ (see Fig.~\ref{Fig2}(c)), which clearly presents trivial and non-trivial phases that are separated by lines $|w|=|v|$. In addition, the charge density distribution for case $w=1.0$, $v=0.5$, $N=40$ (i.e., the red dot in Fig.~\ref{Fig2}(c)) is plot in Fig.~\ref{Fig2}(d), indicating the validity of conventional BBC (see Sec. VIII in Supplementary Material for more details). Note that the quantized NH polarization is protected by CS$^{\dagger}$, though this model also obeys to TRS. The latter is responsible for polarization quantization only when an imaginary-line gap is present (see Table.~\ref{Table1}). Let us stress that in finite lattices, we compute the polarization in terms of real-valued Wannier centers. To do so, singular value decomposition \cite{souza2001maximally} is used to construct numerical Wilson-loop operators (see Sec. IV in the Supplementary Materials for more details).\\
\begin{figure}[t]
    \centering
    \includegraphics[width=0.45\textwidth]{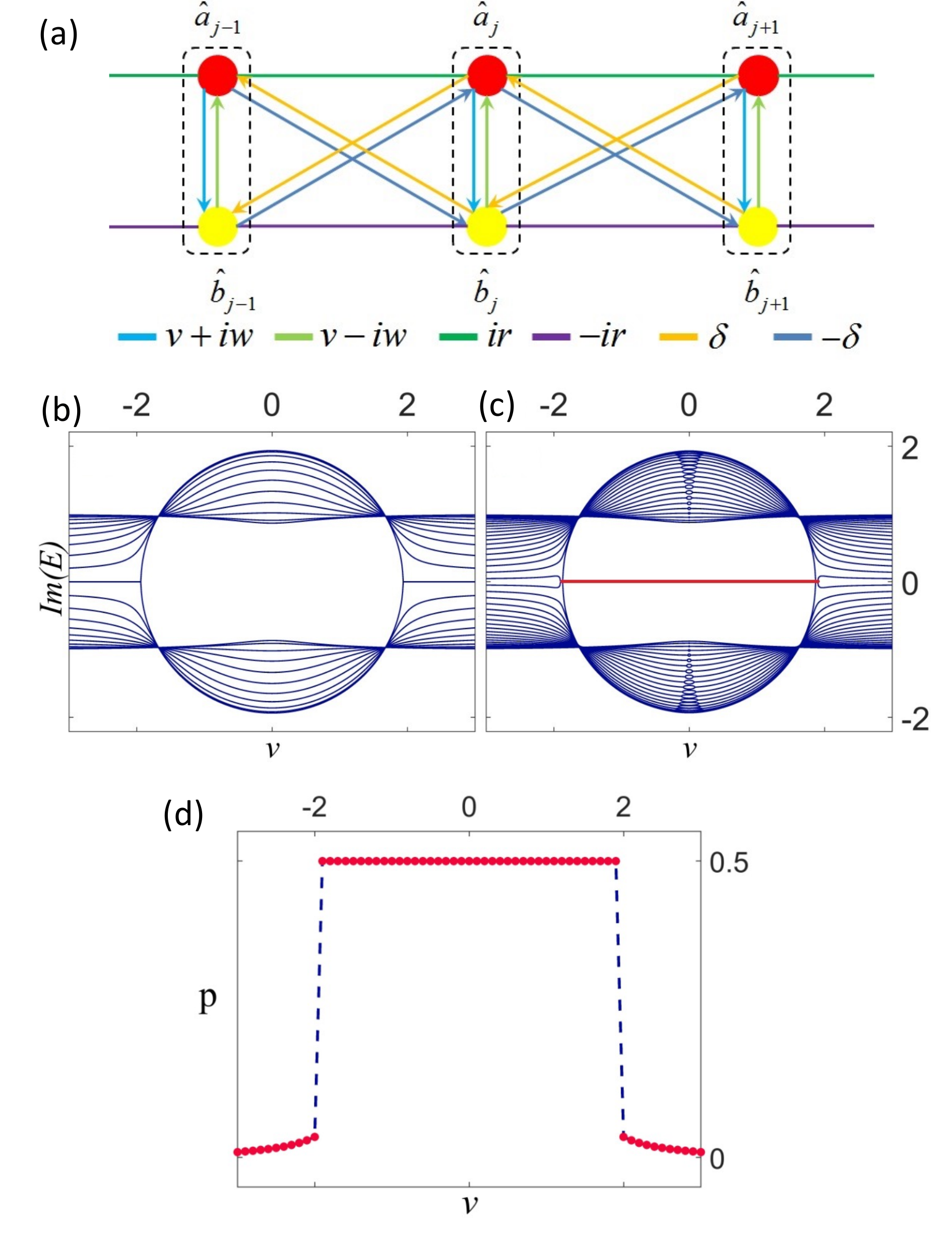}
    \caption{NH polarization and TRS symmetry. (a) Schematic of a one-dimensional NH lattice which obeys only to TRS. The dashed boxes indicate the unit cells. (b,c) Associated imaginary energy spectra, when $w=0.5$, $r=1.0$, $\delta=0.5$, $N=40$, and $v\in(-3,3)$ (see Eq.\ \ref{Eq6}), computed for (a) periodic and (b) open boundary conditions, respectively. The red line in (b) marks the zero-energy edge modes. The energy spectrum presents an imaginary energy gap for $|v|\leq\sqrt{4r^2-w^2}$. (d) Numerical values of the polarization $p$ as a function of $v$, for the same parameters as in panels (b,c).}
    \label{Fig3}
\end{figure}
\indent The second model is schematically shown in Fig.~\ref{Fig3}(a), whose Hamiltonian in momentum-space reads
\begin{equation}\label{Eq6}
    \setlength{\abovedisplayskip}{3pt}
    \setlength{\belowdisplayskip}{3pt}
    H(k)=(v+2i\delta \sin(k))\sigma_x+w\sigma_y+2i r \cos(k)\sigma_z,
\end{equation}
where $(\sigma_x, \sigma_y, \sigma_z)$ are Pauli matrices. It is easy to verify that the Hamiltonian in Eq.~\ref{Eq6} obeys only to TRS with $T=\sigma_x$, and the energy spectrum presents imaginary-line gap for $|v|\leq\sqrt{4r^2-w^2}$, as shown in Fig.~\ref{Fig3}(b,c), while for $|v|\geq\sqrt{4r^2-w^2}$ there is a real-line gap (see Sec. IX in Supplementary Material for complete spectrum). As such, we expect to observe quantized polarization $p=1/2$ in the former case only, as confirmed in numerical simulations shown in Fig.~\ref{Fig3}(d).\\
\noindent \emph{Conclusions} $-$ In summary, we proposed a complete theory for the electric polarization in 1D NH systems presenting line-gapped spectra, and obtained the associated quantization conditions, which are found to be more than those of the Hermitian case. This is essentially related to the possibility that the gap type of line-gapped NH systems may be real or imaginary. We plan to extend these studies to high-order topological multipole moments \cite{benalcazar2017quantized,bradlyn2017topological,schindler2018higher,imhof2018topolectrical,serra2018observation,peterson2018quantized,xue2019acoustic,wang2021quantum,luo2021observation}, whose definition in NH systems look feasible by using the same approach presented here. Although the present paper focuses on the case of independent electrons, Resta's formula was devised for many particle interacting systems \cite{resta1998quantum}, so another potential follow-up of our work is to calculate the electric polarization of NH systems, where many-body effects are included.\\
\begin{acknowledgements}
\noindent We appreciate fruitful discussions with Dr. K. Kawabata, prof. Z. Wang, prof. S. Longhi, Dr. A. Dauphin, prof. P. Massignan, Prof. E. J.  Bergholtz and Prof. M. Nakamura. J. Hu thanks financial support from the National Natural Science Foundation of China (No. 61805141), J. Hu, F. Cardano and L. Marrucci acknowledge financial support from the European Union Horizon 2020 program, under European Research Council (ERC) Grant No. 694683 (PHOSPhOR).
\end{acknowledgements}
\bibliography{Manuscript}

\end{document}


\beginsupplement

\preprint{APS/123-QED}

\title{Supplementary Materials for ‘Electric polarization and its quantization in one-dimensional non-Hermitian chains’}

\author{Jinbing Hu}
\email{hujinbing@usst.edu.cn}
 \affiliation{College of Optical-Electrical Information and Computer Engineering, University of Shanghai for Science and Technology, Shanghai 200093, China.}
 \affiliation{Dipartimento di Fisica “Ettore Pancini”, Università degli Studi di Napoli Federico II, Complesso Universitario di Monte Sant’Angelo, Via Cintia, 80126 Napoli, Italy.}
 \author{Carmine Antonio Perroni}\email{carmine.perroni@unina.it}
\affiliation{Dipartimento di Fisica “Ettore Pancini”, Università degli Studi di Napoli Federico II, Complesso Universitario di Monte Sant’Angelo, Via Cintia, 80126 Napoli, Italy.}
\author{Giulio De Filippis}
\affiliation{Dipartimento di Fisica “Ettore Pancini”, Università degli Studi di Napoli Federico II, Complesso Universitario di Monte Sant’Angelo, Via Cintia, 80126 Napoli, Italy.}
\author{Songlin Zhuang}
\affiliation{College of Optical-Electrical Information and Computer Engineering, University of Shanghai for Science and Technology, Shanghai 200093, China.}
\author{Lorenzo Marrucci}
\affiliation{Dipartimento di Fisica “Ettore Pancini”, Università degli Studi di Napoli Federico II, Complesso Universitario di Monte Sant’Angelo, Via Cintia, 80126 Napoli, Italy.}
\author{Filippo Cardano}
\affiliation{Dipartimento di Fisica “Ettore Pancini”, Università degli Studi di Napoli Federico II, Complesso Universitario di Monte Sant’Angelo, Via Cintia, 80126 Napoli, Italy.}

\maketitle


\section{Unitary position operator}
\noindent Consider a 1D non-Hermitian crystalline system composed of $N$ unit cells, each made of $N_{orb}$ orbitals, the dipole moment density per unit length is  
\begin{equation}\label{S1}
    p=\frac{1}{Na}\left(\sum_{j=1}^{N}\sum_{\alpha=1}^{N_{orb}}q_{j,\alpha}X_{j,\alpha}+\sum_{j=1}^{N}\sum_{\alpha=1}^{N_{orb}}-ex_{j,\alpha}\right), \tag{S1}
\end{equation}
where $X_{j,\alpha}$ are the positions of positive charges $q_{j,\alpha}$, $x_{j,\alpha}$ are the electronic positions, $e$ is the electron charge and $a$ is the unit cell length, which will be set $a=1$ for simplicity unless otherwise specified. The bulk polarization of large crystalline insulator can be equivalently obtained by imposing to the insulator periodic boundary condition (PBC), then performing thermodynamic limit $N\to \infty$. In fact, the contributions of positive charges to the dipole moment can be cancelled out by choosing the center of positive charges (i.e., the atomic nuclei) as the origin of coordinate frame \cite{benalcazar2017electric}, then, the polarization depends only on the positions of electrons. Under PBC, however, the electronic position operator $\hat{x}=\sum_{j=1}^{N}\sum_{\alpha=1}^{N_{orb}}x_{j,\alpha}$ is not a legitimate operator as it does not respect PBC. To overcome this issue, the unitary position operator $\hat{x}_e=exp(-i2\pi\hat{x}/N)$ can be defined \cite{resta1998quantum,benalcazar2017electric}, i.e.,
\begin{equation}\label{S2}
    \hat{x}_e=\sum_{j=1}^{N}\sum_{\alpha=1}^{N_{orb}}c_{j,\alpha}^{\dagger}|0\rangle^{-i\Delta{k}(j+x_{\alpha})}\langle0|c_{j,\alpha}, \tag{S2}
\end{equation}
where $c_{j,\alpha}^{\dagger}(c_{j,\alpha})$ is the creation (annihilation) operator at $\alpha$-orbital of $j$-th cell, $|0\rangle$ is the state of electron, $\Delta{k}=2\pi/N$, and $x_{\alpha}$ is the position of orbital $\alpha$ with respect to the center of positive charge within the unit cell. Using the following discrete Fourier transformation \cite{benalcazar2017electric}
\begin{equation}\label{S3}
    c_{j,\alpha}=\frac{1}{\sqrt{N}}\sum_{k}e^{-ik(j+x_{\alpha})}c_{k,\alpha},\tag{S3}
\end{equation}
where $k\in\Delta{k}\cdot(0,1,\cdots,N-1)$, one has
\begin{align}\label{S4}
    \hat{x}_e&=\sum_{\alpha=1}^{N_{orb}}\sum_{k,k'}\frac{1}{N}\sum_{j=1}^{N}e^{i(k'-k-\Delta{k})(j+x_{\alpha})}c_{k',\alpha}^{\dagger}|0\rangle\langle0|c_{k,\alpha}\notag\\
    &=\sum_{k,\alpha}c_{k+\Delta{k},\alpha}^{\dagger}|0\rangle\langle0|c_{k,\alpha}.\tag{S4}
\end{align}
Here the following identity
$\frac{1}{N}\sum_{j=1}^{N}e^{i(k'-k-\Delta{k})(j+x_{\alpha})}=\delta_{k',k+\Delta{k}}$ is used.\\

\section{Non-Hermitian Projected position operator}
\noindent Regarding above 1D non-Hermitian model, without considering the interaction between electrons, the second-quantized Hamiltonian can be written as \cite{alexandradinata2014wilson}
\begin{equation}\label{S5}
    H=\sum_{k}c_{k,\alpha}^{\dagger}[h_k]^{\alpha\beta}c_{k,\beta},\tag{S5}
\end{equation}
where $\alpha,\beta\in1,2,\dots,N_{orb}$ and summation is implied over repeated orbital indices. In the presence of PBC, there is the following relation
\begin{equation}\label{S6}
    c_{k+Q,\alpha}=e^{iQx_{\alpha}}c_{k,\alpha},\tag{S6}
\end{equation}
where Q is a reciprocal lattice vector. Due to the non-Hermiticity of Hamiltonian, the right and left eigenvectors of k-space Hamiltonian satisfy the following eigenvalue equations
\begin{equation}\label{S7}
    h_k|u_{n,k}^R\rangle=E_{n,k}|u_{n,k}^R\rangle,~h_k^{\dagger}|u_{n,k}^L\rangle=E_{n,k}^*|u_{n,k}^L\rangle,\tag{S7}
\end{equation}
respectively. According to biorthogonal quantum mechanics \cite{brody2013biorthogonal}, if the eigenvalues are linearly independent, the right and left eigenvectors satisfy the biorthogonal relation: $\langle u_{m,k}^L|u_{n,k}^R\rangle=0(m\neq{n})$, and form a complete set of basis. Note that the inner product of right and left eigenvectors $<u_{n,k}^{L}|u_{n,k}^{R}>$ is normally not unit. Therefore, the k-space Hamiltonian can be diagonalized as
\begin{equation}\label{S8}
    [h_k]^{\alpha\beta}=\sum_{n}\frac{[u_{n,k}^R]^{\alpha}E_{n,k}[u_{n,k}^L]^{\beta}}{\langle u_{n,k}^L|u_{n,k}^R\rangle},\tag{S8}
\end{equation}
where $[u_{n,k}^{R/L}]^{\alpha}$ is the $\alpha$-th component of the state $|u_{n,k}^{R/L}\rangle$. Due to the periodicity Eq.~\ref{S6}, the k-space Hamiltonian $h_k$ obeys
\begin{equation}\label{S9}
    h_{k+Q}=V^{-1}(Q)h_kV(Q),\tag{S9}
\end{equation}
with 
\begin{equation}\label{S10}
    [V(Q)]^{\alpha\beta}=e^{-iQx_{\alpha}}\delta_{\alpha\beta},\tag{S10}
\end{equation}
and
\begin{equation}\label{S11}
    [u_{n,k+Q}^{L/R}]=[V^{-1}(Q)]^{\alpha\beta}[u_{n,k}^{L/R}]^{\beta}. \tag{S11}
\end{equation}
Substitute Eq.~\ref{S8} into Eq.~\ref{S5}, we have
\begin{equation}\label{S12}
    H=\sum_{n,k}\gamma_{n,k}^RE_{n,k}\gamma_{n,k}^L,\tag{S12}
\end{equation}
With
\begin{equation}\label{S13}
    \gamma_{n,k}^R=\frac{\sum_{\alpha}c_{k,\alpha}^{\dagger}[u_{n,k}^R]^{\alpha}}{\sqrt{\langle u_{n,k}^L|u_{n,k}^R\rangle}},\tag{S13}
\end{equation}
\begin{equation}\label{S14}
    \gamma_{n,k}^L=\frac{\sum_{\alpha}[u_{n,k}^L]^{\alpha}c_{k,\alpha}}{\sqrt{\langle u_{n,k}^L|u_{n,k}^R\rangle}}.\tag{S14}
\end{equation}
\indent Generally, the physical observable operator can be represented in terms of the eigenvectors of Hamiltonian. However, due to the uncertainty of the eigenvector phase, we here prefer to work with the projection operator, which is defined by 
\begin{equation}\label{S15}
    P=\sum_{n=1}^{N_{orb}}\sum_{k}\gamma_{n,k}^R|0\rangle\langle 0|\gamma_{n,k}^L~.\tag{S15}
\end{equation}
\indent Since the bulk polarization is only related to bands that are lower than line gap, we project the position operator $\hat{x}_e$ [Eq.~\ref{S4}] into the subspace of 'occupied' bands, i.e., 
\begin{align}\label{S16}
    P^{occ}\hat{x}_eP^{occ}&=\sum_{n,k}\sum_{n',k'}\gamma_{n,k+\Delta{k}}^R|0\rangle\langle 0|\gamma_{n,k+\Delta{k}}^L\left(\sum_{q,\alpha}c_{q+\Delta{q},\alpha}^{\dagger}|0\rangle\langle 0|c_{q,\alpha}\right)\gamma_{n',k'}^R|0\rangle\langle 0|\gamma_{n',k'}^L\notag\\
    &=\sum_{n,m=1}^{N_{orb}}\sum_{k}\gamma_{m,k+\Delta{k}}^R|0\rangle [G_k]^{mn}\langle 0|\gamma_{n,k}^L~,\tag{S16}
\end{align}
with the core matrix
\begin{equation}\label{S17}
    [G_k]^{mn}=\frac{\langle u_{m,k+\Delta{k}}^L|u_{n,k}^R\rangle}{\sqrt{\langle u_{m,k+\Delta{k}}^L|u_{m,k+\Delta{k}}^R\rangle\langle u_{n,k}^L|u_{n,k}^R\rangle}}.\tag{S17}
\end{equation}
Here, $N_{occ}$ is the number of bands lower than line gap, which is enabled for non-Hermitian systems featured with line gap (see main text). In the derivation, we have used $\langle 0|\gamma_{n,k}^Lc_{q,\alpha}^{\dagger}|0\rangle=\frac{[u_{n,k}^L]^{\alpha}\delta_{k,q}}{\sqrt{\langle u_{n,k}^L|u_{n,k}^R\rangle}}$ and $\langle 0|c_{q,\alpha}\gamma_{n,k}^R|0\rangle=\frac{[u_{n,k}^R]^{\alpha}\delta_{k,q}}{\sqrt{\langle u_{n,k}^L|u_{n,k}^R\rangle}}$.
 
\section{The core Matrix G}
\noindent The feature of projected position operator is closely related to the property of core matrix G. In Hermitian case, the core matrix G is unitary, then the eigenvalues of the projected position operator represent the positions of occupied-band electrons, i.e., wanner centers. This is also true for non-Hermitian insulator if the core matrix G of non-Hermitian projected position operator Eq.~\ref{S15} is unitary. Therefore, what we need to do is to prove the Hermiticity of core matrix G. To show this, we expand the bra-vector $\langle u_{m,k+\Delta{k}}^L|$ and ket-vector $|u_{m,k+\Delta{k}}^R\rangle$. 
\begin{equation}\label{S18}
    \langle u_{m,k+\Delta{k}}^L|=\langle u_{m,k}^L|+\Delta{k}\cdot\partial_{k}\langle u_{m,k}^L|+\dots,\tag{S18}
\end{equation}
\begin{equation}\label{S19}
    |u_{m,k+\Delta{k}}^R\rangle=|u_{m,k}^R\rangle+\Delta{k}\cdot\partial_{k}|u_{m,k}^R\rangle+\dots.\tag{S19}
\end{equation}
Noting the biorthogonal relation of right and left eigenvectors, we have
\begin{equation}\label{S20}
    \langle\partial_{k}u_{m,k}^L|u_{n,k}^R\rangle=-\langle u_{m,k}^L|\partial_{k}u_{n,k}^R\rangle. \tag{S20}
\end{equation}
Substitute Eq.~\ref{S18}, Eq.~\ref{S19} and Eq.~\ref{S20} into Eq.~\ref{S17} and keep only terms linear in $\Delta{k}$,
\begin{equation}\label{S21}
    \langle u_{m,k+\Delta{k}}^L|u_{m,k+\Delta{k}}^R\rangle=\langle u_{m,k}^L|u_{m,k}^R\rangle+\Delta{k}\cdot\langle\partial_{k}u_{m,k}^L|u_{m,k}^R\rangle+\Delta{k}\cdot\langle u_{m,k}^L|\partial_{k}u_{m,k}^R\rangle,\tag{S21}
\end{equation}
\begin{equation}\label{S22}
    \langle u_{m,k+\Delta{k}}^L|u_{n,k}^R\rangle=\langle u_{m,k}^L|u_{n,k}^R\rangle-\Delta{k}\cdot\langle u_{m,k}^L|\partial_{k}u_{n,k}^R\rangle.\tag{S22}
\end{equation}
The last two terms of Eq.~\ref{S21} cancel out, so the element of core matrix G reads
\begin{equation}\label{S23}
    [G_k]^{mn}=\frac{\langle u_{m,k}^L|u_{n,k}^R\rangle-\Delta{k}\cdot\langle u_{m,k}^L|\partial_{k}u_{n,k}^R\rangle}{\sqrt{\langle u_{m,k}^L|u_{m,k}^R\rangle\langle u_{n,k}^L|u_{n,k}^R\rangle}},\tag{S23}
\end{equation}
which is obviously unitary in thermodynamic limit $N\to\infty$, i.e.,$\Delta{k}\to 0$. Namely, the core matrix G is biorthogonally unitary. Here, by biorthogonally, we mean that the core matrix G is constructed in terms of biorthogonal right and left eigenvectors.

\section{The Wilson loop operator}
\noindent Theoretically, by substituting Eq.~\ref{S23} into the projected position operator Eq.~\ref{S16} and expanding this operator, one can obtain the electronic positions for occupied bands by solving the eigenequation of projected position operator
\begin{equation}\label{S24}
    \left(P^{occ}\hat{x}_eP^{occ}\right)|\Psi^{j}\rangle=\varepsilon^{j}|\Psi^{j}\rangle.\tag{S24}
\end{equation}
This is the case for thermodynamic limit $N\to\infty$, which, however, is not applicable for the case of finite unit cell. To get the eigenvalues of Eq.~\ref{S24} for finite N, we construct the replacing matrix $F_k$ by means of the singular value decomposition \cite{souza2001maximally} of $G_k$, i.e., decomposing G=MDV and defining
\begin{equation}\label{S25}
    F=MV,\tag{S25}
\end{equation}
which is equivalent to G in thermodynamic limit. Using Eq.~\ref{S25} in Eq.~\ref{S24}, we have
\begin{equation}\label{S26}
    \left[
    \begin{array}{ccccc}
        0 & 0 & 0 & \cdots & F_{k_{N}} \\
        F_{k_{1}} & 0 & 0 & \cdots & 0 \\
        0 & F_{k_{2}} & 0 & \cdots & 0 \\
        \vdots & \vdots & \vdots & \ddots & \vdots\\
        0 & 0 & 0 & \cdots & 0
    \end{array}
    \right]
     \left[
        \begin{array}{c}
        \nu_{k_{1}}\\
        \nu_{k_{2}}\\
        \nu_{k_{3}}\\
        \vdots\\
        \nu_{k_{N}}
        \end{array}
     \right]^{j}=
     \varepsilon^{j}
     \left[
     \begin{array}{c}
        \nu_{k_{1}}\\
        \nu_{k_{2}}\\
        \nu_{k_{3}}\\
        \vdots\\
        \nu_{k_{N}}
     \end{array}
     \right]^{j}, \tag{S26}
\end{equation}
where $k_{1}=0,k_{2}=\Delta{k},\dots,k_{N}=(N-1)\cdot\Delta{k}$, and $j\in(1,2,\dots,N_{occ})$. By applying above equation repeatedly, one can get the following relation
\begin{equation}\label{S27}
    W_{k_{f}\gets k_{i}}|\nu_{k_{i}}^j\rangle=\left(\varepsilon^j\right)^{(k_{f}-k_{i})/\Delta{k}}|\nu_{k_{i}}^j\rangle,\tag{S27}
\end{equation}
where we used the bra-ket notation $|\nu_{k_{i}}^j\rangle$ for the vector formed by the collection of values $[\nu_{k_{i}}^j]^n$ for $n\in 1,2,\cdots,N_{occ}$. Comparing Eq.~\ref{S27} and Eq.~\ref{S26}, we define the discrete Wilson line as
\begin{equation}\label{S28}
    W_{k_{f}\gets k_{i}}=F_{k_{f}-\Delta{k}}\cdot F_{k_{f}-2\Delta{k}}\cdots F_{k_{i}+\Delta{k}}\cdot F_{k_{i}}.\tag{S28}
\end{equation}
We can see that even for finite cells the Wilson line is unitary, this is essential to extract real-valued Wannier centers\cite{benalcazar2017electric,alexandradinata2014wilson}. For Wilson loop, i.e., a Wilson line that goes across the entire Brillouin zone, one can get the eigenvalue problem
\begin{equation}\label{S29}
    W_{k+2\pi\gets k}|\nu_{k}^j\rangle=\left(\varepsilon^j\right)^N|\nu_{k}^j\rangle, \tag{S29}
\end{equation}
where $k$ is the starting point of Wilson loop. Note that though the eigenvectors of Wilson loop depend on the starting point, the eigenvalues do not. Since the Wilson loop operator is unitary, then, the eigenvalues are just phases; therefore, Eq.~\ref{S29} can be rewritten as 
\begin{equation}\label{S30}
    W_{k+2\pi\gets k}|\nu_{k}^j\rangle=e^{i2\pi\nu^j}|\nu_{k}^j\rangle, \tag{S30}
\end{equation}
By solving Eq.~\ref{S30} one can obtain the electronic positions $\nu^j$, i.e., \emph{Wannier centers}, based on which the electric polarization can be formulated.

\section{Polarization and complex Berry phase}
\noindent It has been shown that the bulk polarization is a manifestation of the Berry phase of wave functions\cite{resta1998quantum,king1993theory}. For instance, in Hermitian insulators the bulk polarization is proportional to Berry phase that the subspace of occupied bands accumulates as it is parallel transported around the Brillouin zone (BZ) \cite{king1993theory,resta1993towards}. Topologically, the Berry phase is a geometric phase of Hilbert space constructed by complete eigenvectors. As mentioned above since the right or left eigenvectors of non-Hermitian Hamiltonian considered is complete, then, a complete Hilbert space can also be built in terms of right and left eigenvectors \cite{shen2018topological}. Parallel transporting the subspace of occupied bands across BZ one cycle also accumulates $2n\pi$ ($n$ is integer). To demonstrate this relation, we consider the original definition of Wilson loop
\begin{equation}\label{S31}
    W_{k+2\pi\gets k}=G_{k+2\pi-\Delta{k}}\cdot G_{k+2\pi-2\Delta{k}}\cdots G_{k+\Delta{k}}\cdot G_{k}.\tag{S31}
\end{equation}
Substitute Eq.~\ref{S23} into above equation Eq.~\ref{S31} and if we define the complex Berry connection \cite{lieu2018topological} of non-Hermitian system as
\begin{equation}\label{S32}
    [A_k]^{mn}=-i\frac{\langle u_{m,k}^L|\partial_{k}u_{n,k}^R\rangle}{\sqrt{\langle u_{m,k}^L|u_{m,k}^R\rangle\langle u_{n,k}^L|u_{n,k}^R\rangle}},\tag{S32}
\end{equation}
then, we can obtain the Wilson loop in thermodynamic limit
\begin{equation}\label{S33}
    W_{k+2\pi\gets k}=exp(-i\int_{k}^{k+2\pi}A_k dk).\tag{S33}
\end{equation}
Since the Wannier centers are the phases of the eigenvalues of the Wilson loop \cite{benalcazar2017electric}, the bulk polarization of non-Hermitian insulator can be alternatively written as
\begin{align}\label{S34}
     p&=-\frac{i}{2\pi}log~ det\left[e^{-i\int_{k}^{k+2\pi}A_kdk}\right]\notag\\
    &=-\frac{1}{2\pi}\oint_{BZ}Tr[A_k]dk,~~mod~1,\tag{S34}
\end{align}
which is similar to the well-known expression for the polarization in the modern theory of polarization\cite{king1993theory,resta1998quantum,benalcazar2017electric}. That is, the bulk polarization of non-Hermitian insulators is the manifestation of complex Berry phase \cite{lieu2018topological} of complete biorthogonal Hilbert space. 
\begin{figure}
    \centering
    \includegraphics[width=0.38\textwidth]{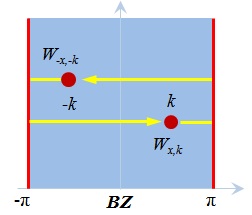}
    \caption{The schematic for the definition of Wilson loop. Since  $W_{x,k}$ and $W_{-x,k}$ are along opposite directions, there is such a relation: $W_{-x,k}=W_{x,k}^{\dagger}$. Note that the eigenvalues of Wilson loop operator depends not on the base point $k$. }
    \label{SMFig1}
\end{figure}

\section{The constraint of symmetry to Wilson-loop operator}
\noindent From above Sec. III, we know that the Wilson loop operator is biorthogonally unitary and its eigenvalues still represent the Wannier centers, i.e., the relative displacement of electron charge center with respect to the atomic center. Thus, we can determine the quantization of 1D non-Hermitian polarization imposed by a specific symmetry by analyzing the restriction of the specific symmetry to Wilson-loop operator. Before proceeding, we first define the notation of Wilson-loop operator. We denote the Wilson-loop operator with base point $k$ that progresses towards larger value until reaching $k+2\pi$ as
\begin{equation}\label{S35}
    W_{x,k}\equiv G_{k+N\Delta{k}}\cdot G_{k+(N-1)\Delta{k}}\cdots G_{k+\Delta{k}}\cdot G_{k}. \tag{S35}
\end{equation}
Similarly, we denote the Wilson-loop operator with base point $k$ advancing towards smaller value until reaching $k-2\pi$ as
\begin{equation}\label{S36}
    W_{-x,k}\equiv G_{k-N\Delta{k}}\cdot G_{k-(N-1)\Delta{k}}\cdots G_{k-\Delta{k}}\cdot G_{k}. \tag{S36}
\end{equation}
In addition, these Wilson-loop operators obey [Fig.~\ref{SMFig1}]
\begin{equation}\label{S37}
    W_{-x,k}=W_{x,k}^{\dagger}.\tag{S37}
\end{equation}
Here, we would like to stress that the element of Wilson-loop operator has the form
\begin{equation}\label{S38}
    W_{k}^{mn}\equiv\langle U_{m}^L(k)|U_{n}^R(k)\rangle, \tag{S38}
\end{equation}
with $|U_{m}^i(k)\rangle=|u_{m}^{i}(k)\rangle/\sqrt{\langle u_{m}^{L}(k)|u_{m}^{R}(k)\rangle}(i=R,L)$, which is subject to the completeness condition
\begin{equation}\label{S39}
    \sum_{n}|U_{n}^R(k)\rangle\langle U_{n}^L(k)|=\sum_{n}|U_{n}^L(k)\rangle\langle U_{n}^R(k)|=1.\tag{S39}
\end{equation}
For convenience, we will omit the summation symbol in the following. \\
\indent In non-Hermitian physics, internal symmetries ramify due to the distinction of complex conjugation and transposition \cite{kawabata2019symmetry}. The conventional time-reversal symmetry (TRS) is divided into two kinds; one is normal TRS defined in terms of complex conjugation, i.e., $Th^*(k)T^{-1}=h(-k)$, the other is anomalous TRS (TRS$^{\dagger}$) defined in terms of transposition, i.e., $\tilde{T}h^t(k)\tilde{T}^{-1}=h(-k)$. Similarly, chiral symmetry (CS) is ramified into normal CS, satisfying $\Gamma h^{\dagger}(k)\Gamma^{-1}=-h(k)$ and anomalous CS (CS$^{\dagger}$) defined as $Sh(k)S=-h(k)$, particle-hole symmetry (PHS) is split into normal PHS, satisfying $Ch^t(k)C^{-1}=-h(-k)$and anomalous PHS (PHS$^{\dagger}$) defined as $\tilde{C}h^*(k)\tilde{C}^{-1}=-h(-k)$. In addition, there is another basic symmetry, inversion symmetry $I$ (IS). Therefore, there are seven basic symmetries in total. Note that all above symmetry operators are unitary. In what follows, we analyze the restriction imposed by seven basic symmetries to the Wilson-loop operator.

\subsection{Chiral symmetry}
A non-Hermitian insulator with chiral symmetry obeys\cite{kawabata2019symmetry}
\begin{equation}\label{S40}
    \Gamma h^{\dagger}(k)\Gamma^{-1}=-h(k),\tag{S40}
\end{equation}
where $\Gamma$ is the unitary chiral operator, satisfying $\Gamma^2=1$. By means of Eq.~\ref{S40}, we can get
\begin{equation}\label{S41}
    h^{\dagger}(k)\Gamma|U_{n}^R(k)\rangle=-E_n(k)\Gamma|U_{n}^R(k)\rangle,\tag{S41}
\end{equation}
which implies that $\Gamma|U_{n}^R(k)\rangle$ is a left eigenstate with corresponding energy $-E_n(k)$. Since $h(k)$ has a complete basis, therefore, one can expand $\Gamma|U_{n}^R(k)\rangle$ in terms of the basis of $h(k)$:
\begin{equation}\label{S42}
    \Gamma|U_{n}^R(k)\rangle=|U_{m}^L(k)\rangle\langle U_{m}^R(k)|\Gamma|U_{n}^R(k)\rangle=|U_{m}^L(k)\rangle B_{\Gamma,k}^{mn},\tag{S42}
\end{equation}
with $B_{\Gamma,k}^{mn}=\langle U_{m}^R(k)|\Gamma|U_{n}^R(k)\rangle$, and summation over repeated band indices has been implied. Since $\Gamma|U_{n}^R(k)\rangle$ is a left eigenstate with corresponding energy $-E_n(k)$, then,$B_{\Gamma,k}^{mn}$ connects the states at k point on opposite sides of the energy gap, i.e., $E_n(k)=-E_m(k)$. Here, we can always, at least locally, choose a gauge of the biorthogonal basis such that $B_{\Gamma,k}^{mn}$ is a unitary matrix independent of $k$ \cite{kawabata2019symmetry}. On the basis of Eq.~\ref{S42}, we can get
\begin{equation}\label{S43}
    |U_{n}^R(k)\rangle=\Gamma^{\dagger}|U_{m}^L(k)\rangle B_{\Gamma,k}^{mn},\tag{S43}
\end{equation}
and
\begin{equation}\label{S44}
    \langle U_{n}^L(k)|=B_{\Gamma,k}^{\dagger^{nm}}\langle U_{m}^R(k)|\Gamma.\tag{S44}
\end{equation}
Using Eqs.~\ref{S43} and \ref{S44} the element of a Wilson line from $k_1$ to $k_2$ reads
\begin{align}\label{S45}
     W_{k_2\gets K_1}^{mn}&=\langle U_{m}^L(k_2)|U_{n}^R(k_1)\rangle\notag\\
    &=B_{\Gamma,k_2}^{\dagger^{ms}}\langle U_{s}^R(k_2)|\Gamma\Gamma^{\dagger}|U_{r}^L(k_1)\rangle B_{\Gamma,k_1}^{rn}\notag\\
    &=B_{\Gamma,k_2}^{\dagger^{ms}}W_{k_2\gets k_1}^{sr}B_{\Gamma,k_1}^{rn}.\tag{S45}
\end{align}
Since $B_{\Gamma,k}^{mn}$ connects the states at k point on opposite sides of the energy gap, Eq.~\ref{S45} represents the connection of Wilson lines between occupied and unoccupied bands. In particular, for a Wilson loop at base point $k$ we have
\begin{equation}\label{S46}
    W_{k}^{unocc}=B_{\Gamma,k}^{\dagger}W_{k}^{occ}B_{\Gamma,k},\tag{S46}
\end{equation}
where occ (unocc) stands for Wilson loop over occupied (unoccupied) bands. Equation~\ref{S46} means that the Wannier centers from the occupied bands are equal to those from the unoccupied bands. Note that the Hilbert space over all bands, including occupied and unoccupied bands, is topologically trivial, thus, the bulk polarization from both occupied and unoccupied bands is necessarily trivial\cite{benalcazar2017electric}. Therefore, the polarization is quantized to be either 0 or 1/2 in the presence of chiral symmetry.

\subsection{Anomalous chiral symmetry}
A non-Hermitian insulator with anomalous chiral symmetry obeys\cite{kawabata2019symmetry}
\begin{equation}\label{S47}
    Sh(k)S^{-1}=-h(k),\tag{S47}
\end{equation}
where $S$ is the unitary chiral operator, satisfying $S^2=1$. By means of Eq.~\ref{S47}, we can get
\begin{equation}\label{S48}
    h(k)S|U_{n}^R(k)\rangle=-E_n(k)S|U_{n}^R(k)\rangle,\tag{S48}
\end{equation}
implying that if $|U_{n}^R(k)\rangle$ is a eigenstate of occupied bands with corresponding eigenenergy $E_n(k)$, then, $S|U_{n}^R(k)\rangle$ must be a eigenstate of unoccupied bands with corresponding energy $-E_n(k)$. Same as before, one can expand $S|U_{n}^R(k)\rangle$ in terms of the basis of $h(k)$:
\begin{equation}\label{S49}
    S|U_{n}^R(k)\rangle=|U_{m}^L(k)\rangle\langle U_{m}^R(k)|S|U_{n}^R(k)\rangle=|U_{m}^L(k)\rangle B_{S,k}^{mn},\tag{S49}
\end{equation}
with $B_{S,k}^{mn}=\langle U_{m}^R(k)|S|U_{n}^R(k)\rangle$ that connects the states at k point on opposite sides of the energy gap, i.e., $E_n(k)=-E_m(k)$. The following procedure is similar as that of chiral symmetry. At last, we can reach the following relation of Wilson loops between occupied and unoccupied bands 
\begin{equation}\label{S50}
    W_{k}^{unocc}=B_{S,k}^{\dagger}W_{k}^{occ}B_{S,k}.\tag{S50}
\end{equation}
Same as the case of chiral symmetry, Eq.~\ref{S50} implies that the non-Hermitian polarization is restricted to be either 0 or 1/2 in the presence of anomalous chiral symmetry.

\subsection{Inversion symmetry}
A non-Hermitian insulator with inversion symmetry obeys\cite{kawabata2019symmetry}
\begin{equation}\label{S51}
    Ih(k)I^{-1}=h(-k),\tag{S51}
\end{equation}
where $I$ is the unitary inversion operator, satisfying$II^{\dagger}=I^{\dagger}I=1$. The state $I|U_{n}^R(k)\rangle$ is an eigenstate of $h(-k)$ with energy as can be shown as follows:
\begin{equation}\label{S52}
    h(-k)I|U_{n}^R(k)\rangle=E_n(k)I|U_{n}^R(k)\rangle,\tag{S52}
\end{equation}
Since the eigenstates of $h(-k)$ form a compete basis, one can expand $I|U_{n}^R(k)\rangle$ in terms of the basis of $h(-k)$:
\begin{equation}\label{S53}
    I|U_{n}^R(k)\rangle=|U_{m}^R(-k)\rangle\langle U_{m}^L(-k)|I|U_{n}^R(k)\rangle=|U_{m}^R(-k)\rangle B_{I,k}^{mn},\tag{S53}
\end{equation}
with $B_{I,k}^{mn}=\langle U_{m}^L(-k)|I|U_{n}^R(k)\rangle$. On the basis of Eq.~\ref{S53} we can get
\begin{equation}\label{S54}
    |U_{n}^R(k)\rangle=I^{\dagger}|U_{m}^R(-k)\rangle B_{I,k}^{mn},\tag{S54}
\end{equation}
and
\begin{equation}\label{S55}
    \langle U_{n}^L(k)|=B_{I,k}^{\dagger^{nm}}\langle U_{m}^L(-k)|I,\tag{S55}
\end{equation}
Therefore, an element of a Wilson line from $k_1$ to $k_2$ reads
\begin{align}\label{S56}
    W_{k_2\gets k_1}^{mn}&=\langle U_{m}^L(k_2)|U_{n}^R(k_1)\rangle\notag\\
    &=B_{I,k_2}^{\dagger^{ms}}\langle U_{s}^L(-k_2)|II^{\dagger}|U_{r}^R(-k_1)\rangle B_{I,k_1}^{rn}\notag\\
    &=B_{I,k_2}^{\dagger^{ms}}W_{-k_2\gets -k_1}^{sr}B_{I,k_1}^{rn}.\tag{S56}
\end{align}
In particular, for a Wilson loop with base point $k$ we have
\begin{equation}\label{S57}
    W_{x,k}=B_{I,k}^{\dagger}W_{x,-k}^{\dagger}B_{I,k},\tag{S57}
\end{equation}
in which we have used Eq.~\ref{S37}. Since the eigenvalues of Wilson-loop operator is independent of base point, Eq.~\ref{S57} implies that the set of Wilson-loop eigenvalues have to be equal to its complex conjugate. Consequently, the polarization of 1D non-Hermitian insulator is quantized to be either 0 or 1/2 by inversion symmetry.

\subsection{Time-reversal symmetry}
A non-Hermitian insulator respecting time-reversal symmetry has\cite{kawabata2019symmetry}
\begin{equation}\label{S58}
    Th^*(k)T^{-1}=h(-k),\tag{S58}
\end{equation}
where $T$ is the unitary time-reversal operator, satisfying $TT^{\dagger}=T^{\dagger}T=1$. Perform Hermitian conjugation to both sides of Eq.~\ref{S58}, we can get
\begin{equation}\label{S59}
    h^{\dagger}(-k)=Th^t(k)T^{-1},\tag{S59}
\end{equation}
where the superscript $t$ denotes transposition. By means of Eq.~\ref{S59}, it is straightforward to show that $\langle U_{n}^R(k)|T^{-1}$ gives a left eigenstate with eigenenergy $E_n(k)$. Similar as before, we can expand $\langle U_{n}^R(k)|T^{-1}$ in terms of the basis of $h(-k)$:
\begin{equation}\label{S60}
    \langle U_{n}^R(k)|T^{-1}=\langle U_{n}^R(k)|T^{-1}|U_{m}^R(-k)\rangle\langle U_{m}^L(-k)|=B_{T,k}^{nm}\langle U_{m}^L(-k)|,\tag{S60}
\end{equation}
with $B_{T,k}^{nm}=\langle U_{n}^R(k)|T^{-1}|U_{m}^R(-k)\rangle$. On the basis of Eq.~\ref{S60} we can get
\begin{equation}\label{S61}
    \langle U_{n}^R(k)|=B_{T,k}^{nm}\langle U_{m}^L(-k)|T,\tag{S61}
\end{equation}
and
\begin{equation}\label{S62}
    |U_{n}^L(k)\rangle=T^{\dagger}|U_{m}^R(-k)\rangle B_{T,k}^{\dagger^{mn}},\tag{S62}
\end{equation}
Using Eqs.~\ref{S61} and \ref{S62}, the element of a Wilson line from $k_1$ to $k_2$ is equal to
\begin{align}\label{S63}
    W_{k_2\gets k_1}^{mn}&=\langle U_{m}^R(k_2)|U_{n}^L(k_1)\rangle\notag\\
    &=B_{T,k}^{ms}\langle U_{s}^L(-k_2)|TT^{\dagger}|U_{r}^R(-k_1)\rangle B_{T,k}^{\dagger^{rn}}\notag\\
    &=B_{T,k}^{ms}W_{-k_2\gets -k_1}^{sr}B_{T,k}^{\dagger^{rn}}.\tag{S63}
\end{align}
In particular, for a Wilson loop with base point $k$ we have
\begin{equation}\label{S64}
    W_{x,k}=B_{T,k}W_{x,-k}^{\dagger}B_{T,k}^{\dagger},\tag{S64}
\end{equation}
Similarly, Eq.~\ref{S64} implies that the set of Wilson-loop eigenvalues have to be equal to its complex conjugate. Consequently, the polarization of 1D non-Hermitian insulator is quantized to be either 0 or 1/2 by time-reversal symmetry. However, one critical attention should be paid, that is, above conclusion only applies to time-reversal symmetry with $T^*T=1$, because for $T^*T=-1$ the sewing matrix is zero\cite{kawabata2019symmetry}. The reason is as follow.
According to the unitary feature of time-reversal operator, we have
\begin{equation}\label{S65}
    TT^{\dagger}=1\to T^*T^{t}=1,\tag{S65}
\end{equation}
Considering $T^*T=-1$, we can get $T^*T^t=-T^*T$, from which $T^t=-T$. Due to this equality, we also have
  \begin{equation}\label{S66}
      \langle U_{m}^R(-k)|T^{-1}|U_{n}^R(k)\rangle=\langle U_{m}^R(-k)|T^{t^{-1}}|U_{n}^R(k)\rangle=-\langle U_{m}^R(-k)|T^{-1}|U_{n}^R(k)\rangle,\tag{S66}
  \end{equation}
which leads to $\langle U_{m}^R(-k)|T^{-1}|U_{n}^R(k)\rangle=0$.

\subsection{Anomalous time-reversal symmetry}
A non-Hermitian insulator respecting anomalous time-reversal symmetry obeys\cite{kawabata2019symmetry}
\begin{equation}\label{S67}
    \tilde{T}h^t(k)\tilde{T}^{-1}=h(-k),\tag{S67}
\end{equation}
where $\tilde{T}$ is the unitary anomalous time-reversal operator, satisfying $\tilde{T}\tilde{T}^{\dagger}=\tilde{T}^{\dagger}\tilde{T}=1$. Perform Hermitian conjugation to both sides of Eq.~\ref{S67}, we can get
\begin{equation}\label{S68}
    h^{\dagger}(-k)=\tilde{T}h^*(k)\tilde{T}^{-1},\tag{S68}
\end{equation}
which leads to
  \begin{equation}\label{S69}
      h^{\dagger}(k)\tilde{T}|U_{n}^{R^*}(-k)\rangle=\tilde{T}h^*(-k)|U_{n}^{R^*}(-k)\rangle=E_{n}^*(-k)\tilde{T}|U_{n}^{R^*}(-k)\rangle.\tag{S69}
  \end{equation}
Equation~\ref{S69} implies that $\tilde{T}|U_{n}^{R^*}(-k)\rangle$ gives a left eigenstate with corresponding eigenenergy$E_{n}^*(-k)$, which can be expanded in terms of the basis of $h(k)$:
\begin{equation}\label{S70}
    \tilde{T}|U_{n}^{R^*}(-k)\rangle=|U_{m}^L(k)\rangle\langle U_{m}^R(k)|\tilde{T}|U_{n}^{R^*}(-k)\rangle=|U_{m}^L(k)\rangle B_{\tilde{T},k}^{mn},\tag{S70}
\end{equation}
with $B_{\tilde{T},k}^{mn}=\langle U_{m}^R(k)|\tilde{T}|U_{n}^{R^*}(-k)\rangle$. On the basis of Eq. ~\ref{S70} we can get
\begin{equation}\label{S71}
    |U_{n}^L(k)\rangle=\tilde{T}|U_{m}^{R^*}(-k)\rangle B_{\tilde{T},k}^{\dagger^{mn}},\tag{S71}
\end{equation}
and
\begin{equation}\label{S72}
    \langle U_{m}^R(k)|=B_{\tilde{T},k}^{mn}\langle U_{n}^{L^*}(-k)|\tilde{T}^{\dagger},\tag{S72}
\end{equation}
On the basis of Eqs.~\ref{S71} and~\ref{S72}, the element of a Wilson line from $k_1$ to $k_2$ reads
\begin{align}\label{S73}
    W_{k_2\gets k_1}^{mn}&=\langle U_{m}^R(k_2)|U_{n}^L(k_1)\rangle\notag\\
    &=B_{\tilde{T},k_2}^{ms}\langle U_{s}^{L^*}(-k_2)|\tilde{T}^{\dagger}\tilde{T}|U_{r}^{R^*}(-k_1)\rangle B_{\tilde{T},k_1}^{\dagger^{rn}}\notag\\
    &=B_{\tilde{T},k_2}^{ms}W_{-k_2\gets -k_1}^{sr}B_{\tilde{T},k_1}^{\dagger^{rn}}. \tag{S73}
\end{align}
In particular, for a Wilson loop with base point $k$ we have
\begin{equation}\label{S74}
    W_{x,k}=B_{\tilde{T},k}W_{x,-k}B_{\tilde{T},k}^{\dagger},\tag{S74}
\end{equation}
in which we have used Eq.~\ref{S37}. Since the eigenvalues of Wilson-loop operator is independent of base point, Eq.~\ref{S74} obviously shows that anomalous time-reversal symmetry impose no constraint to the polarization of 1D non-Hermitian insulator, which is much distinct with the case of time-reversal symmetry.

\subsection{Particle-hole symmetry}
A non-Hermitian insulator with particle-hole symmetry obeys\cite{kawabata2019symmetry}
\begin{equation}\label{S75}
    Ch^t(k)C^{-1}=-h(-k),\tag{S75}
\end{equation}
where $C$ is the unitary anomalous particle-hole operator, subject to $CC^{\dagger}=C^{\dagger}C=1$. Perform Hermitian conjugation to both sides of Eq.~\ref{S75}, we can get
\begin{equation}\label{S76}
    h^{\dagger}(-k)=-Ch^*(k)C^{-1},\tag{S76}
\end{equation}
which leads to
  \begin{equation}\label{S77}
      h^{\dagger}(k)C|U_{n}^{R^*}(-k)\rangle=-Ch^*(-k)|U_{n}^{R^*}(-k)\rangle=-E^*_n(-k)C|U_{n}^{R^*}(-k)\rangle. \tag{S77}
  \end{equation}
Equation \ref{S75} implies that $C|U_{n}^{R^*}(-k)\rangle$ gives a left eigenstate with eigenenergy $-E_{n}^*(-k)$. Similar as before, we can expand $C|U_{n}^{R^*}(-k)\rangle$ in terms of the basis of $h(k)$:
\begin{equation}\label{S78}
    C|U_{n}^{R^*}(-k)\rangle=|U_{m}^L(k)\rangle\langle U_{m}^R(k)|C|U_{n}^{R^*}(-k)\rangle=|U_{m}^L(k)\rangle B_{C,k}^{mn},\tag{S78}
\end{equation}
with $B_{C,k}^{mn}=\langle U_{m}^R(k)|C|U_{n}^{R^*}(-k)\rangle$. On the basis of Eq.~\ref{S78} we can get
\begin{equation}\label{S79}
    |U_{n}^L(k)\rangle=C|U_{m}^{R^*}(-k)\rangle B_{C,k}^{\dagger^{mn}},\tag{S79}
\end{equation}
and
\begin{equation}\label{S80}
    \langle U_{n}^R(k)|=B_{C,k}^{nm}\langle U_{m}^{L^*}(-k)|C^{\dagger},\tag{S80}
\end{equation}
By means of above two equations, the element of a Wilson line from $k_1$ to $k_2$ is
\begin{align}\label{S81}
    W_{k_2\gets k_1}^{mn}&=\langle U_{m}^R(k_2)|U_{n}^L(k_1)\rangle\notag\\
    &=B_{C,k_2}^{ms}\langle U_{s}^{L^*}(-k_2)|C^{\dagger}C|U_{r}^{R^*}(-k_1)\rangle B_{C,k_1}^{\dagger^{rn}}\notag\\
    &=B_{C,k_2}^{ms}W_{-k_2\gets -k_1}^{*^{sr}}B_{C,k_1}^{\dagger^{rn}}.\tag{S81}
\end{align}
In particular, for a Wilson loop with base point $k$ we have
\begin{equation}\label{S82}
    W_{k}^{occ}=B_{C,k}W_{-k}^{unocc}B_{C,k}^{\dagger},\tag{S82}
\end{equation}
in which we have used Eq.~\ref{S37}. Note that different from the sewing matrix of the case of anomalous time-reversal symmetry, i.e., Eq.~\ref{S74}, the sewing matrix here connects states at $k$ with states at $-k$ such that $E_m(k)=-E_n(-k)$. Thus, for a nonzero $B_{C,k}^{mn}$, if $m$ labels a state in the occupied bands, then, $n$ labels a state in the unoccupied band, or vice versa. Therefore, same as the case of chiral symmetry (Eq.~\ref{S46}) and anomalous chiral symmetry (Eq.~\ref{S50}), Eq.~\ref{S82} implies the quantization of 1D non-Hermitian polarization by particle-hole symmetry. Here again, we stress that this polarization quantization is applicable to non-Hermitian insulator subject to particle-hole symmetry with $CC^*=1$, for $CC^*=-1$, the sewing matrix is zero due to the same reason as that of TRS (see subsection. 4. Time-reversal symmetry). 

\subsection{Anomalous particle-hole symmetry}
A non-Hermitian insulator with anomalous particle-hole symmetry obeys\cite{kawabata2019symmetry}
\begin{equation}\label{S83}
    \tilde{C}h^*(k)\tilde{C}^{-1}=-h(-k),\tag{S83}
\end{equation}
where $\tilde{C}$ is the unitary anomalous particle-hole operator, satisfying $\tilde{C}\tilde{C}^{\dagger}=\tilde{C}^{\dagger}\tilde{C}=1$. Perform Hermitian conjugation to both sides of Eq.~\ref{S83}, we can get
\begin{equation}\label{S84}
    h^{\dagger}(-k)=-\tilde{C}h^t(k)\tilde{C}^{-1},\tag{S84}
\end{equation}
which leads to
  \begin{equation}\label{S85}
      \langle U_{n}^R(k)|\tilde{C}^{-1}h^{\dagger}(-k)=-\langle U_{n}^R(k)|h^t(k)\tilde{C}^{-1}=-\langle U_{n}^R(k)|\tilde{C}^{-1}E_n(k).\tag{S85}
  \end{equation}
Equation~\ref{S85} implies that $\langle U_{n}^R(k)|\tilde{C}^{-1}$ is a left eigenstate with eigenenergy $-E_n(k)$, which can be expanded in terms of the basis of $h(-k)$:
  \begin{equation}\label{S86}
    \langle U_{n}^R(k)|\tilde{C}^{-1}=\langle U_{n}^R(k)|\tilde{C}^{-1}|U_{m}^R(-k)\rangle\langle U_{m}^L(-k)|=B_{\tilde{C},k}^{nm}\langle U_{m}^L(-k)|.\tag{S86}
\end{equation}
with $B_{\tilde{C},k}^{nm}=\langle U_{n}^R(k)|\tilde{C}^{-1}|U_{m}^R(-k)\rangle$. On the basis of Eq. (S86) we can get
\begin{equation}\label{S87}
    \langle U_{n}^R(k)|=B_{\tilde{C},k}^{nm}\langle U_{m}^L(-k)|\tilde{C},\tag{S87}
\end{equation}
and
\begin{equation}\label{S88}
    |U_{n}^L(k)\rangle=\tilde{C}^{\dagger}|U_{m}^R(-k)\rangle B_{\tilde{C},k}^{\dagger^{mn}}.\tag{S88}
\end{equation}
Therefore, by means of Eqs.~\ref{S88} and \ref{S88}, the element of a Wilson line from $k_1$ to $k_2$ is equal to
\begin{align}\label{S89}
    W_{k_2\gets k_1}^{mn}&=\langle U_{m}^R(k_2)|U_{n}^L(k_1)\rangle\nonumber\\
    &=B_{\tilde{C},k_2}^{ms}\langle U_{s}^L(-k_2)|\tilde{C}\tilde{C}^{\dagger}|U_{r}^R(-k_1)\rangle B_{\tilde{C},k_1}^{\dagger^{rn}}\nonumber\\
    &=B_{\tilde{C},k_2}^{ms}W_{-k_2\gets -k_1}^{sr}B_{\tilde{C},k_1}^{\dagger^{rn}}.\tag{S89}
\end{align}
In particular, for a Wilson loop with base point $k$ we have
\begin{equation}\label{S90}
    W_{k}^{occ}=B_{\tilde{C},k}W_{-k}^{unocc}B_{\tilde{C},k}^{\dagger},\tag{S90}
\end{equation}
Note that the physical meaning of sewing matrix here is same as that in particle-hole symmetry. Once again, Eq.~\ref{S90} is only applicable to non-Hermitian insulators subject to anomalous particle-hole symmetry with $\tilde{C}\tilde{C}^*=1$. For $\tilde{C}\tilde{C}^*=-1$, the sewing matrix is zero (see Subsec. 4 Time-reversal symmetry). That is, the bulk polarization of 1D non-Hermitian insulator is quantized by anomalous particle-hole symmetry to be either 0 or 1/2 only when $\tilde{C}\tilde{C}^*=1$.

\section{Symmetries and delocalized bulk states}
\noindent If the wave number $k$ of bulk wave function is complex, then, the wave function is exponentially decaying at the rate of imaginary part of wave number. So to form electric polarization, the bulk-mode wave functions should be extended. According to Ref.\cite{yi2020non}, by solving characteristic equation $f(\beta,E)=0$ the energy spectrum for long chain lattice and the profile of generalized Brillouin zone (GBZ) of non-Hermitian systems can be obtained. Since $f(\beta,E)=0$, we also have $[f(\beta,E)]^*=0$. In what follows, we reveal the features of energy spectrum and of bulk states that above seven basic symmetries impose.

\subsection{Time-reversal symmetry}
For TRS, the Bloch Hamiltonian under periodic boundary condition satisfies\cite{yi2020non}
\begin{equation}\label{S91}
    Th^*(k)T^{-1}=h(-k),\tag{S91}
\end{equation}
which implies the characteristic equation 
\begin{equation}\label{S92}
    f(k,E)=det[E-h(k)]=\left(det[E^*-h(-k)]\right)^*=f^*(-k,E^*),\tag{S92}
\end{equation}
Now we extend to the complex plane by using the relation $\beta=e^{ik}$, the characteristic equation becomes
\begin{equation}\label{S93}
    f(\beta,E)=f^*(\beta^*,E^*),\tag{S93}
\end{equation}
Since $f(\beta,E)=\left[f(\beta,E)\right]^*=0$, then, we have
\begin{equation}\label{S94}
    f(\beta,E)=f(\beta^*,E^*),\tag{S94}
\end{equation}
Equation~\ref{S94} implies that the energies of system with TRS come in pairs $(E,E^*)$, i.e., possessing imaginary line gap.

\subsection{Anomalous time-reversal symmetry}
For TRS$^{\dagger}$, the Bloch Hamiltonian with periodic boundary condition satisfies\cite{yi2020non}
\begin{equation}\label{S95}
    \tilde{T}h^t(k)\tilde{T}^{-1}=h(-k),\tag{S95}
\end{equation}
which implies the characteristic equation 
\begin{equation}\label{S96}
    f(k,E)=det[e-h(k)]=det[E-h^t(-k)]=f(-k,E),\tag{S96}
\end{equation}
Similarly, by expanding to complex plane, the characteristic equation becomes
\begin{equation}\label{S97}
    f(\beta,E)=f(1/\beta,E),\tag{S97}
\end{equation}
Equation~\ref{S97} implies that in the presence of TRS$^{\dagger}$, a system presents delocalized bulk states \cite{kawabata2019symmetry,zhang2020correspondence,okuma2020topological}.

\subsection{Chiral symmetry}
For CS, the Bloch Hamiltonian with periodic boundary condition satisfies\cite{yi2020non}
\begin{equation}\label{S98}
    \Gamma h^{\dagger}(k)\Gamma=-h(k),\tag{S98}
\end{equation}
which implies the characteristic equation 
\begin{equation}\label{S99}
    f(k,E)=det[E-h(k)]=(-1)^d\left(det[-E^*-h^t(k)]\right)^*=(-1)^df^*(k,-E^*),\tag{S99}
\end{equation}
where $d$ is the dimension of Hamiltonian matrix, which is normally even number. Same as before, by expanding to complex plane, the characteristic equation becomes
\begin{equation}\label{S100}
    f(\beta,E)=f(1/\beta^*,-E^*),\tag{S100}
\end{equation}
Above equation implies that system CS presents real line energy gap\cite{kawabata2019symmetry} and extended bulk states \cite{kawabata2019symmetry,zhang2020correspondence,okuma2020topological}.

\subsection{Anomalous chiral symmetry}
For CS$^{\dagger}$, the Bloch Hamiltonian with periodic boundary condition satisfies\cite{kawabata2019symmetry}
\begin{equation}\label{S101}
    Sh(k)S=-h(k),\tag{S101}
\end{equation}
which implies the characteristic equation 
\begin{equation}\label{S102}
    f(k,E)=det[E-h(k)]=(-1)^d det[-E-h(k)]=(-1)^d f(k,-E),\tag{S102}
\end{equation}
where $d$ is the dimension of Hamiltonian matrix and is normally even number. By expanding to complex plane, the characteristic equation becomes
\begin{equation}\label{S103}
    f(\beta,E)=f(\beta,-E),\tag{S103}
\end{equation}
That is, symmetry CS$^{\dagger}$ imposes no restriction to the system energy spectrum and bulk states \cite{kawabata2019symmetry,zhang2020correspondence,okuma2020topological}.

\subsection{Inversion symmetry}
For system with IS, the Bloch Hamiltonian with periodic boundary condition satisfies
\begin{equation}\label{S104}
    Ih(k)I^{-1}=h(-k),\tag{S104}
\end{equation}
which implies the characteristic equation 
\begin{equation}\label{S105}
    f(k,E)=det[E-h(k)]=det[E-h(-k)]=f(-k,E),\tag{S105}
\end{equation}
By expanding to complex plane, the characteristic equation becomes
\begin{equation}\label{S106}
    f(\beta,E)=f(1/\beta,E),\tag{S106}
\end{equation}
That is, a system with IS exhibits delocalized bulk states \cite{kawabata2019symmetry,zhang2020correspondence,okuma2020topological}.

\subsection{Particle-hole symmetry}
For PHS, the Bloch Hamiltonian with periodic boundary condition obeys\cite{kawabata2019symmetry}
\begin{equation}\label{S107}
    Ch^t(k)C^{-1}=-h(-k),\tag{S107}
\end{equation}
which implies the characteristic equation 
\begin{equation}\label{S108}
    f(k,E)=det[E-h(k)]=(-1)^d det[-E-h^t(-k)]=(-1)^d f(-k,-E),\tag{S108}
\end{equation}
where $d$ is the dimension of Hamiltonian matrix, which is normally even number. Same as before, by expanding to complex plane, the characteristic equation becomes
\begin{equation}\label{S109}
    f(\beta,E)=f(1/\beta,-E),\tag{S109}
\end{equation}
Therefore, a system with PHS presents extended bulk states \cite{kawabata2019symmetry,zhang2020correspondence,okuma2020topological}.

\subsection{Anomalous particle-hole symmetry}
For PHS$^{\dagger}$, the Bloch Hamiltonian with periodic boundary condition obeys\cite{kawabata2019symmetry}                     
\begin{equation}\label{S110}
    \tilde{C}h^*(k)\tilde{C}^{-1}=-h(-k),\tag{S110}
\end{equation}
which implies the characteristic equation 
\begin{equation}\label{S111}
    f(k,E)=det[E-h(k)]=(-1)^d\left(det[-E^*-h(-k)]\right)^*=(-1)^d f^*(-k,-E^*),\tag{S111}
\end{equation}
where $d$ is the dimension of Hamiltonian matrix, which is normally even number. Same as before, by expanding to complex plane, the characteristic equation becomes
\begin{equation}\label{S112}
    f(\beta,E)=f^*(1/\beta,-E^*),\tag{S112}
\end{equation}
Therefore, a system with PHS$^{\dagger}$ presents extended bulk states \cite{kawabata2019symmetry,zhang2020correspondence,okuma2020topological} and real line gap\cite{kawabata2019symmetry}.
\begin{figure}
    \centering
    \includegraphics[width=0.7\textwidth]{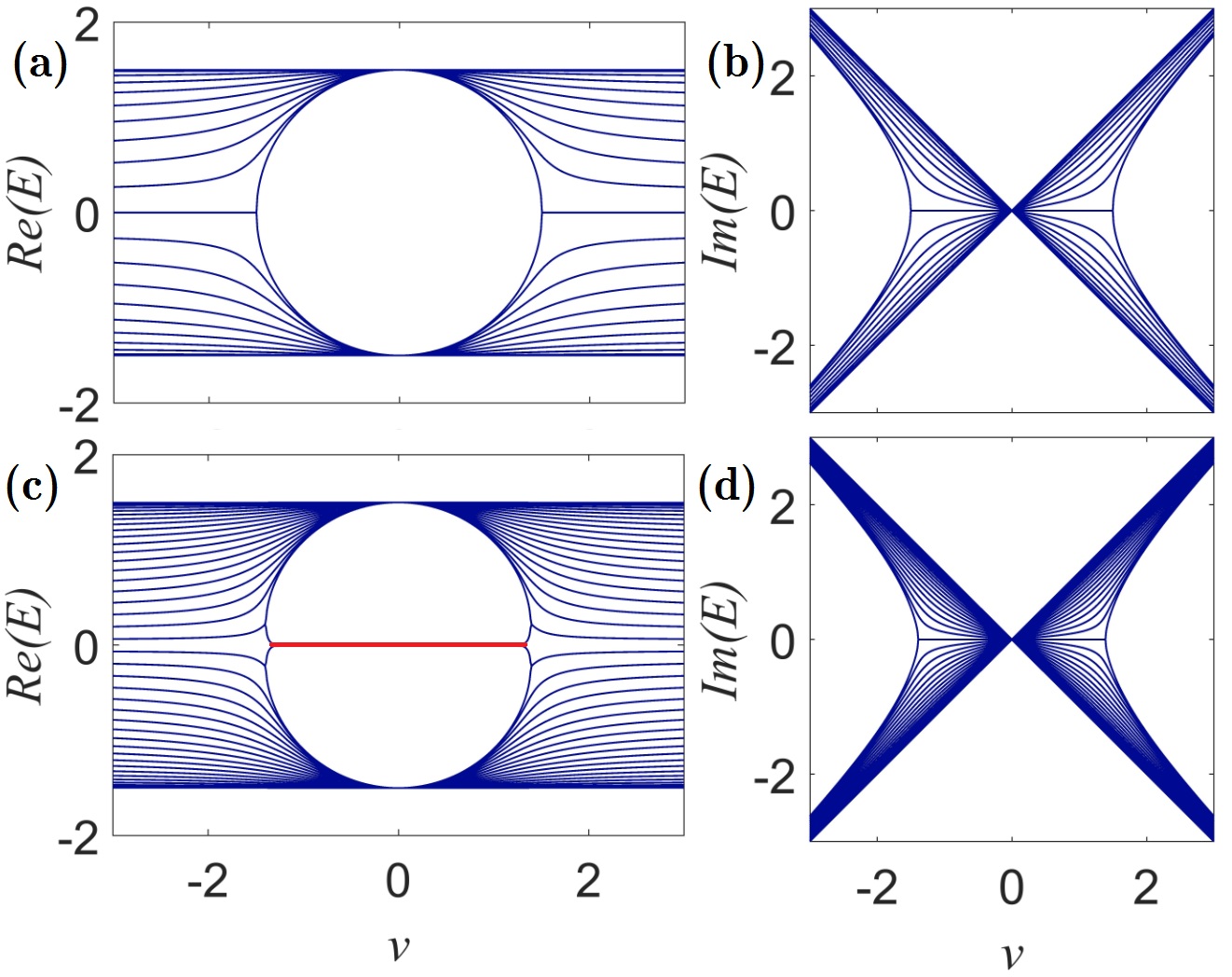}
    \caption{The real and imaginary parts of energy spectra of the non-Hermitian Hamiltonian Eq.~\ref{S114} with $w = 1.5$,$v\in (-3,3)$, and $N=40$. (a) Real and (b) Imaginary parts of the spectra for periodic boundary condition (PBC), and (c) and (d) for open boundary condition (OBC). Red line in (c) denote zero-energy edge modes. }
    \label{SMFig2}
\end{figure}
\begin{figure}
    \centering
    \includegraphics[width=0.8\textwidth]{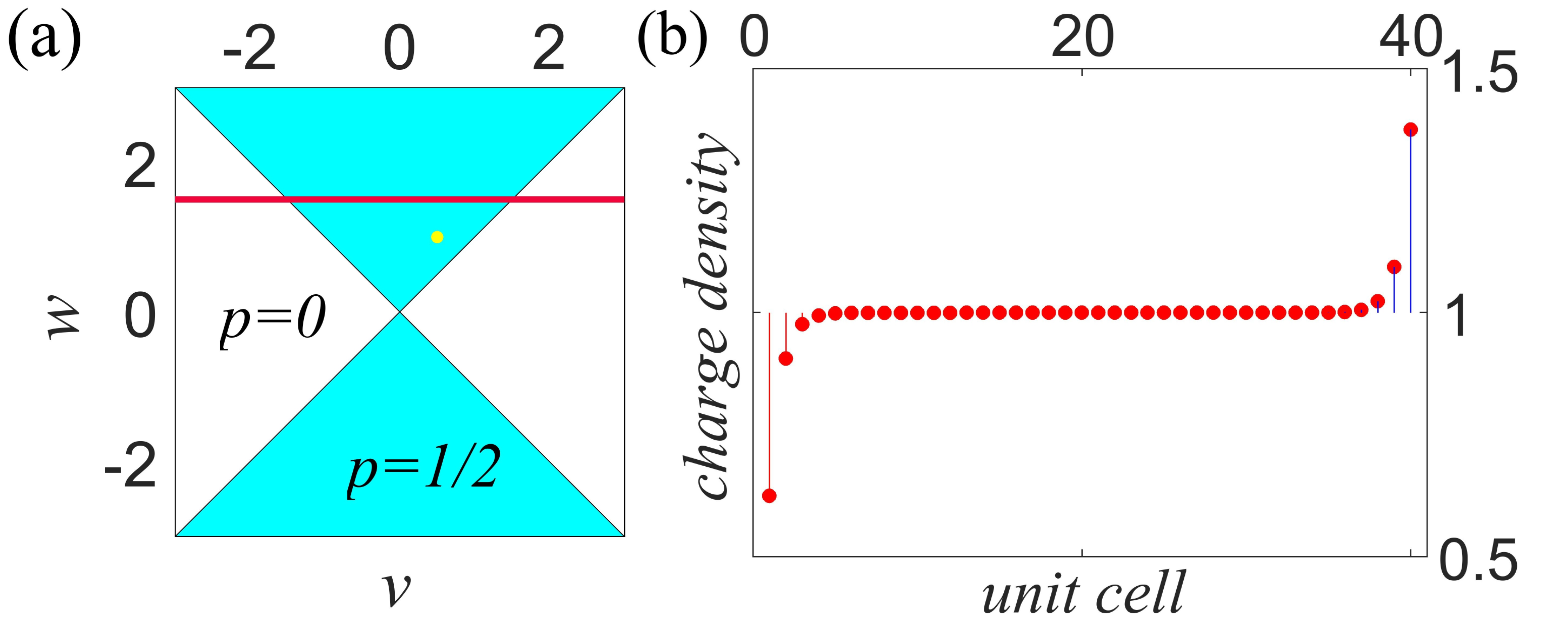}
    \caption{(a) the polarization phase diagram as functions of intercell hopping $w$ and intracell hopping $v$. The red line corresponds to the situation plot in Fig. \ref{SMFig2}, and the yellow dot denotes the case plot in (b). (b) Electron charge distribution in the topologically nontrivial polarization phase $(w = 1.0,v = 0.5)$. The total electronic charge at the ends is $\pm e/2$ relative to background.}
    \label{SMFig3}
\end{figure}
\section{Hamiltonian with real-line gap}
\noindent To convince above theory, we here present a non-Hermitian model presenting polarization in the presence of real-line gap. Without loss of generality, we consider the following 1D two-sites model, in which we assume symmetric intercell hoppings. The Hamiltonian reads
\begin{equation}\label{S113}
    \hat{H}=\sum_{j}\left[v(\hat{a}_{j}^{\dagger}\hat{b}_{j}-\hat{b}_{j}^{\dagger}\hat{a}_{j})+w(\hat{a}_{j}^{\dagger}\hat{b}_{j-1}+\hat{b}_{j-1}^{\dagger}\hat{a}_{j})\right],\tag{S113}
\end{equation}
where $\hat{a}_j (\hat{a}_j^{\dagger})$ and $\hat{b}_j (\hat{b}_j^{\dagger})$ are annihilation (creation) operators on the sites ‘A’ and ‘B’ of $j$-th cell, respectively, and $v,w\in\mathbb{R}$ denote hopping amplitudes. The minus sign means that the forward and backward hoppings inside the unit cell differ by $\pi$, which can be experimentally implemented \cite{hafezi2011robust,hafezi2013imaging,mittal2019photonic}. In the presence of translational invariance, the Hamiltonian can be conveniently expressed in the space of quasi-momentum  
\begin{equation}\label{S114}
    h(k)=d_x(k)\sigma_x+(d_y(k)+iv)\sigma_y,\tag{S114}
\end{equation}
where $\sigma_i(i=x,y)$ is Pauli matrix, and $d_x(k)=w\cdot cos(k)$,  $d_y(k)=w\cdot sin(k)$. Obviously, the k-space Hamiltonian respects CS$^{\dagger}$ and TRS with $S=\sigma_z$ and $T=\sigma_0$ respectively, which force the eigenenergies of Eq.~\ref{S114} to come in $(E, -E^*)$ pair:  
\begin{equation}\label{S115}
    E_{\pm}=\pm\sqrt{w^2-v^2+2\, i\,wv\, sin(k)},\tag{S115}
\end{equation}
\indent Figure~\ref{SMFig2} shows the real and imaginary parts of energy spectra for both periodic ((a) and (b)) and open ((c) and (d)) boundary conditions for $w = 1.5,~v\in(-3,3),~N = 40$. We can see that although the energy spectrum is complex for $v\in(-3,3)$ due to the presence of non-Hermiticity, it presents real-line gap, as shown by real parts of spectra, and this is guaranteed by CS$^{\dagger}$ and TRS. Thus, at half-filling, the model is an insulator. In addition, by comparison we find that the bulk energy spectra remain unchanged when boundary conditions alter from periodic to open, implying that the wave functions of bulk states are delocalized\cite{kawabata2019symmetry,yi2020non}. In terms of the theory of non-Hermitian electric polarization i.e., Eq.~3 in the main text, we calculate and plot the topological phase diagram as functions of $w$ and $v$, as shown in Fig.~\ref{SMFig3}(a), which shows that the model is in topologically nontrivial polarization phase when $|w|>|v|$, i.e., $p=1/2$; while it is in topologically trivial phase when $|w|<|v|$, i.e., $p=0$. Bulk polarization manifests itself through the appearance of charge at its ends \cite{vanderbilt1993electric}, so we also plot the charge density distribution of the case $w=1.0,~v=0.5,~N=40$ (i.e., the yellow dot in Fig.~\ref{SMFig3}(a)) in Fig.~\ref{SMFig3}(b), which evidently shows the charge deviation at both ends of the chain from one electron per cell, which is the case of PBC when the model is at half-filling.
\begin{figure}
    \centering
    \includegraphics[width=0.6\textwidth]{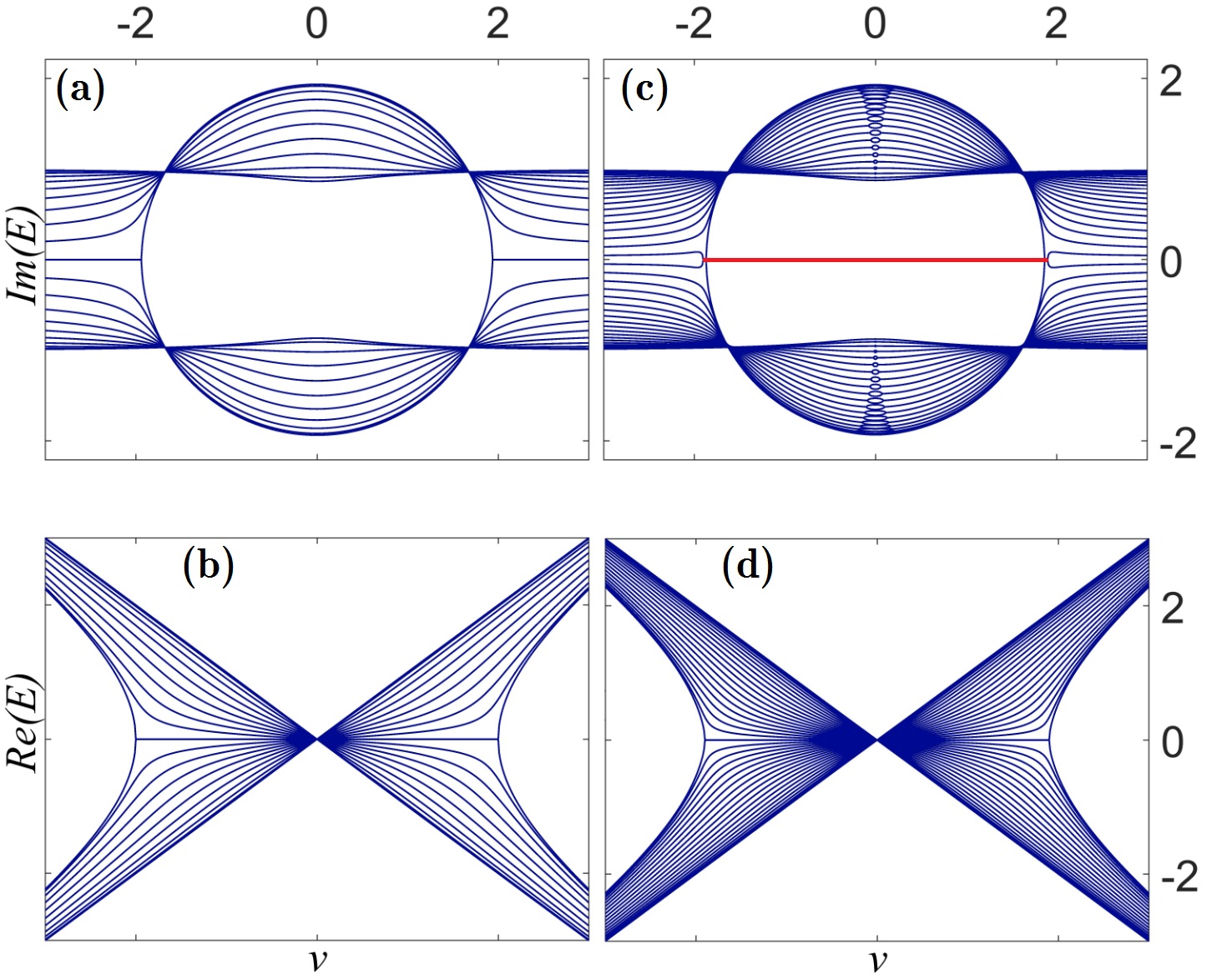}
    \caption{The energy spectra of non-Hermitian Hamiltonian Eq.~\ref{S117} for $w=0.5$, $r=1.0$, $\delta=0.5$, $N=40$, and $v\in(-3,3)$. (a) and (b) [(c) and (d)] are the imaginary and real parts of energies, respectively, for periodic [open] boundary conditions. Red line in (c) denotes zero-energy edge modes. The imaginary energy spectrum closes at points $v=\pm\sqrt{4r^2-w^2}$, implying phase transition at these points.} 
    \label{SMFig4}
\end{figure}
\begin{figure}
    \centering
    \includegraphics[width=0.5\textwidth]{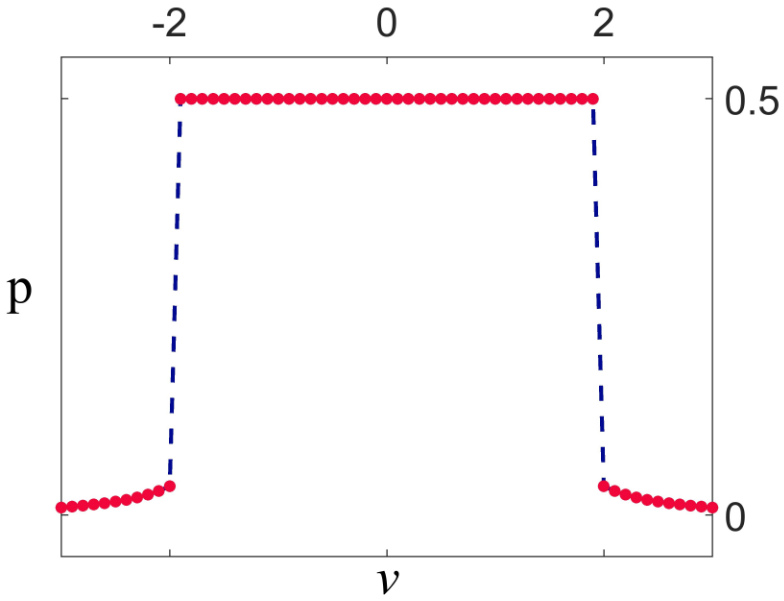}
    \caption{Numerical result of topological invariant $\nu$ as a function of $v$ for $w=0.5$, $r=1.0$, $\delta=0.5$. Note that the phase transition points locate $v=\pm1.93$} 
    \label{SMFig5}
\end{figure}
\begin{figure}
    \centering
    \includegraphics[width=0.7\textwidth]{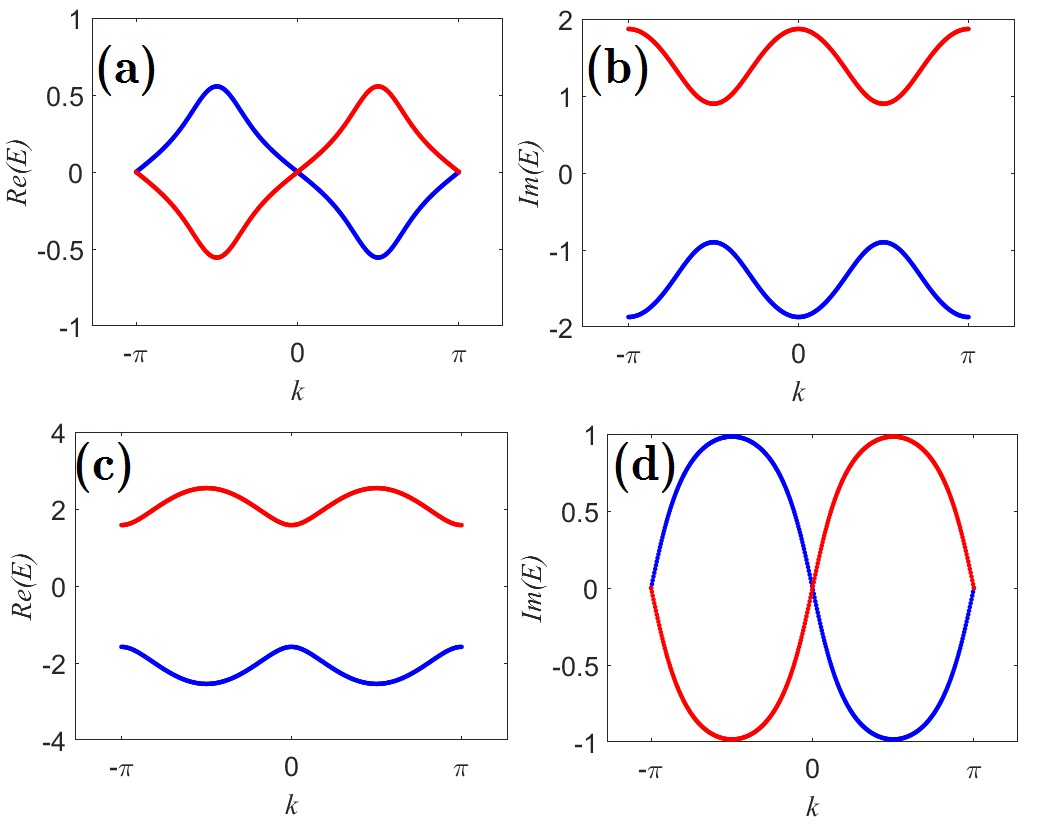}
    \caption{Real and imaginary energy spectra of Eq.~\ref{S118} for two sets of parameters: (a) and (b) $v=0.5$, $w=0.5$, $r=1.0$ and $\delta=0.5$, (c) and (d) $v=2.5$, $w=0.5$, $r=1.0$ and $\delta=0.5$. The energy spectrum of the former set of parameters presents imaginary-line gap (see (b)) while that of the latter set of parameters exhibits real-line gap (see (c))..} 
    \label{SMFig6}
\end{figure}
\section{Hamiltonian with imaginary-line gap}
In this section, we consider another non-Hermitian model that respects only time-reversal symmetry. The schematic is shown in Fig.~4 of main text, 1D non-Hermitian tight-binding model composed by periodically separated bi-particle unit cells with Hamiltonian being of the form
\begin{align}\label{S116}
    \hat{H}=&\sum_j{[(v-iw)a_j^{\dagger}b_j+(v+iw)b_j^{\dagger}a_j]}+\sum_j{[ira_j^{\dagger}a_{j+1}+ira_{j+1}^{\dagger}a_j]}\notag\\
    &-\sum_j{[irb_j^{\dagger}b_{j+1}+irb_{j+1}^{\dagger}b_j]}+\sum_j{[\delta b_{j-1}^{\dagger}a_j+\delta a_{j-1}^{\dagger}b_j]}-\sum_j{[\delta a_j^{\dagger}b_{j-1}+\delta b_j^{\dagger}a_{j-1}]},\tag{S116}
\end{align}
where $a_j^{\dagger}$ ($b_j^{\dagger}$) and $a_j$ ($b_j$) are the creation and annihilation operators at $a$-site ($b$-site) of $j$-th unit cell, and these hoppings $v,w,r,\delta\in\mathbb{R}$. By means of Fourier transformation, the momentum-space Hamiltonian can be written as 
\begin{equation}\label{S117}
    H(k)=(v+2i\delta sin(k))\sigma_x+w\sigma_y+2i r cos(k)\sigma_z.\tag{S117}
\end{equation}
It is easy to verify that the Hamiltonian Eq.~\ref{S117} respects only TRS in the form $TH^*(k)T^{-1}=H(-k)$ \cite{kawabata2019symmetry} with $T=\sigma_x$, where $*$ denotes complex conjugation. By diagonalizing Hamiltonian matrix, one can get the eigenenergies
\begin{equation}\label{S118}
    E(k)=\pm\sqrt{(v+2i\delta sin(k))^2+w^2-4r^2cos^2(k)}.\tag{S118}
\end{equation}
\indent Figure ~\ref{SMFig4} shows the energy spectra for a set of parameters with periodic [Fig.~\ref{SMFig4}(a) and (b)] and open [Fig.~\ref{SMFig4}(c) and (d)] boundary conditions (BCs). It is obviously that the imaginary part of $E(k)$ is gaped when $|v|\leq\sqrt{4r^2-w^2}$ due to the presence of only TRS. According to Table. I of main text, in the presence of imaginary-line gap, TRS quantizes non-Hermitian polarization. By means of Eq. 3 of the main text, the non-Hermitian polarization is computed, and the results are plot in Fig.~\ref{SMFig5}. However, we note that for $|v|\geq\sqrt{4r^2-w^2}$ the bulk polarization is neither 0 nor 1/2, this is because in that region the model Eq.~\ref{S118} exhibits real-line gap, which is clearly shown in Fig.~\ref{SMFig6}. \\

\bibliography{SMReference}